\documentclass[aps,prb,reprint,groupedaddress]{revtex4-2}
\usepackage{amsmath}
\usepackage{amssymb}
\usepackage{graphicx}
\usepackage{subfigure}
\usepackage{comment}
\begin{document}
   \title{Duality between Generalized Non-Hermitian Hatano-Nelson Model in Flat Space and Hermitian System in Curved Space}
   \author{Shu-Xuan Wang }
   \email{wshx123@mail.ustc.edu.cn}
   \affiliation{Department of Modern Physics, University of Science and Technology of China, Hefei, 230026, China}
   \author{Shaolong Wan}
   \email{slwan@ustc.edu.cn}
    \affiliation{Department of Modern Physics, University of Science and Technology of China, Hefei, 230026, China}
   \date{\today}
   \begin{abstract}
    In this paper, we explore a possibility for the non-Hermiticity of the system originating from the curved space. We obtain a duality between a generalized non-Hermitian Hatano-Nelson (HN) model in $d$-dimensional flat space and a Hermitian system in $2d$-dimensional curved space, and give the metric of the curved space analytically. This duality shows a possibility of non-Hermiticity coming from metric of the space, and provides a new perspective on exploring non-Hermitian or Hermitian systems.
   \end{abstract}
   \maketitle

   \section{Introduction}
       In the past, the properties of Hermitian systems is main topic in condensed matter physics. In recent years, non-Hermitian systems have been widely investigated and many novel phenomena have been explored\cite{15,16,17,18,19,20,21,22,24,25,26,27,28,29,30,32,33,34,35,36,37,38}. Different from Hermitian systems, there are numerous localized states  at the boundaries in non-Hermitian systems, which is called skin effect\cite{17,27,28,21}. The traditional Bloch band theory is invalid in non-Hermitian systems. Hence, non-Bloch band theory is proposed to depict the complex spectra and skin effect in non-Hermitian systems\cite{19,26}. However, the non-Bloch band theory is only valid for $1$-dimensional non-Hermitian systems. Similar to the Hermitian systems, the bulk-boundary correspondence~(BBC) is established based on non-Bloch band theory\cite{18,26}, and the topological invariants are calculated over the generalized Brillouin zones~(GBZs). Moreover, the effective topological field theory has also been developed to elaborate the BBC in non-Hermitian systems\cite{36}. In addition, the existence of exceptional points/rings is also intriguing in non-Hermitian systems, where the Hamiltonian is defective and complex bands coalesce\cite{16,20,32,33,34}.
       \par

       What the non-Hermiticity of the system's Hamiltonian originates from is a fundamental question. In general, for an open quantum system, the non-Hermitian Hamiltonian is an effective Hamiltonian\cite{37,38}, its non-Hermiticity comes from gain and loss of the system. For a closed quantum system, its non-Hermiticity comes from the energy exchange with surroundings. For theoretical self-consistent, when these conditions disappear, the Hamiltonian of the system must return to Hermitian one from non-Hermitian one. We wonder whether other origins of non-Hermiticity of the system exist. Recently, a perspective arouse our interest, which indicate that the non-Hermiticity can relate to curved space even if this system is isolated\cite{30}. It shows geometric root of non-Hermiticity of the systems. So a non-Hermitian system can be explored by studying the corresponding system in curved space if we find the relation between them and vice versa.
       \par

       In this article, we establish a duality between non-Hermitian HN model in $d$-dimensional flat space and Hermitian system in $2d$-dimensional curved space. By comparing Schr\"{o}dinger equations of the two systems, we get the metric of the $2d$-dimensional curved space analytically, which contains $d$ parameters. The results show that the non-Hermiticity of the system can originate from the curved space.
       \par

       This article is organized as follows. In Sec.$\mathbf{II}$, we derive the duality between a non-Hermitian HN model in $1$-dimensional  flat space and a Hermitian system in $2$-dimensional curved space, and obtain the metric of the curved space analytically. In Sec.$\mathbf{III}$, we study high-dimensional cases, and give the duality between non-Hermitian HN model in $d$-dimensional flat space and Hermitian system in $2d$-dimensional curved space. In Sec.$\mathbf{IV}$, we show how to give a non-Hermitian HN model in flat space from the given metric of curved space. In Sec.$\mathbf{V}$, we discuss some special cases. Finally, conclusion and discussion are given in Sec.$\mathbf{VI}$.

   \section{Metric of the curved space corresponding to  1-dimensional non-Hermitian HN model}
       For simplicity, we first consider the non-Hermitian HN model in $1$-dimensional flat space with open boundary condition. The generic Hamiltonian of this model is
          \begin{equation}
            H=\sum_{n,j}t_{j} c_{n+j}^{\dagger} c_n , \label{1}
          \end{equation}
       where, $n\in [1,L]$, is the index of site, and $j\in [j_{min},j_{max}]$, is the index of hopping length. Considering translation symmetry, hopping amplitude $t_j$ is independent of $n$. Due to the absence of magnetic field, $t_j \geqslant 0$ for any $j$.
       \par

       Assume that $\Psi=(\psi_1,\psi_2,\cdots,\psi_L)^T$ is a solution of Schr\"{o}dinger equation of this system with eigenenergy $E$, where $\psi_i$ is the amplitude of the wavefunction on the $i$th site. Then, Schr\"{o}dinger equation can be simplified as
           \begin{equation}
               \sum_j t_j \psi_{n-j} = E \psi_{n} . \label{2}
           \end{equation}
        Considering skin effect\cite{17,19}, we take wavefunction as $\psi_n= e^{-qn}\phi_n$, where $q \in \mathbb{R}$ and $e^{-q}$ is the decay factor. $\Phi=(\phi_1,\phi_2,\cdots,\phi_L)^T$ is the extended part of the wavefunction. Substituting $\psi_n$ into Eq.(2), we obtain
            \begin{equation}
                \sum_j t_j e^{qj} \phi_{n-j} =E \phi_n  . \label{3}
            \end{equation}
        To derive the duality, we utilize the continuous position variable $x$ instead~(function $\phi(x)$ about variable $x$ satisfies $\phi(x)|_{x=n}=\phi_n$ and we set lattice constant $a=1$, $\psi(x)=e^{-qx} \phi(x)$). According to non-Bloch band theory\cite{19}, $\phi(x) = |\phi(x)| e^{-i \theta x}$ and $|\phi(x)|$ changes smoothly with the variation of position. Now, we consider the case $\theta \in [-\theta_1, \theta_2]$, $0 <\theta_1, \theta_2 < \pi$ and $\frac{2\pi}{max\{ \theta_1, \theta_2\} } >> (j_{max}-j_{min})$. Thus, $ e^{-i \theta x}$ changes smoothly with the variation of position under this case. Applying Taylor expansion to $\phi(x)$ at $x=n$ and truncate it at the second order term~(Appendix F), $\phi(x+j) = \phi(x) + j\partial_x \phi(x) + \frac{j^2}{2} \partial_x^2 \phi(x)$. Substituting it into Eq.\eqref{3}, we  get
            \begin{equation}
                B \partial_x^2 \phi(x) + C \partial_x \phi(x) = A \phi(x) ,\label{4}
            \end{equation}
        where $B=\sum_j \frac{j^2}{2} t_j e^{qj} >0$, $C=\sum_j -j t_j e^{qj}$ and $A=E-\sum_j t_j e^{qj}$.

        \par

        Then, we introduce a new coordinate $y=e^{qx}$, and express Eq.(4) as
            \begin{equation}
                \left[B(qy)^2 \partial_y^2 + (Bq^2y + Cqy) \partial_y \right]\phi = A\phi .\label{5}
            \end{equation}
        \par

\begin{figure}
            \includegraphics[scale=0.55]{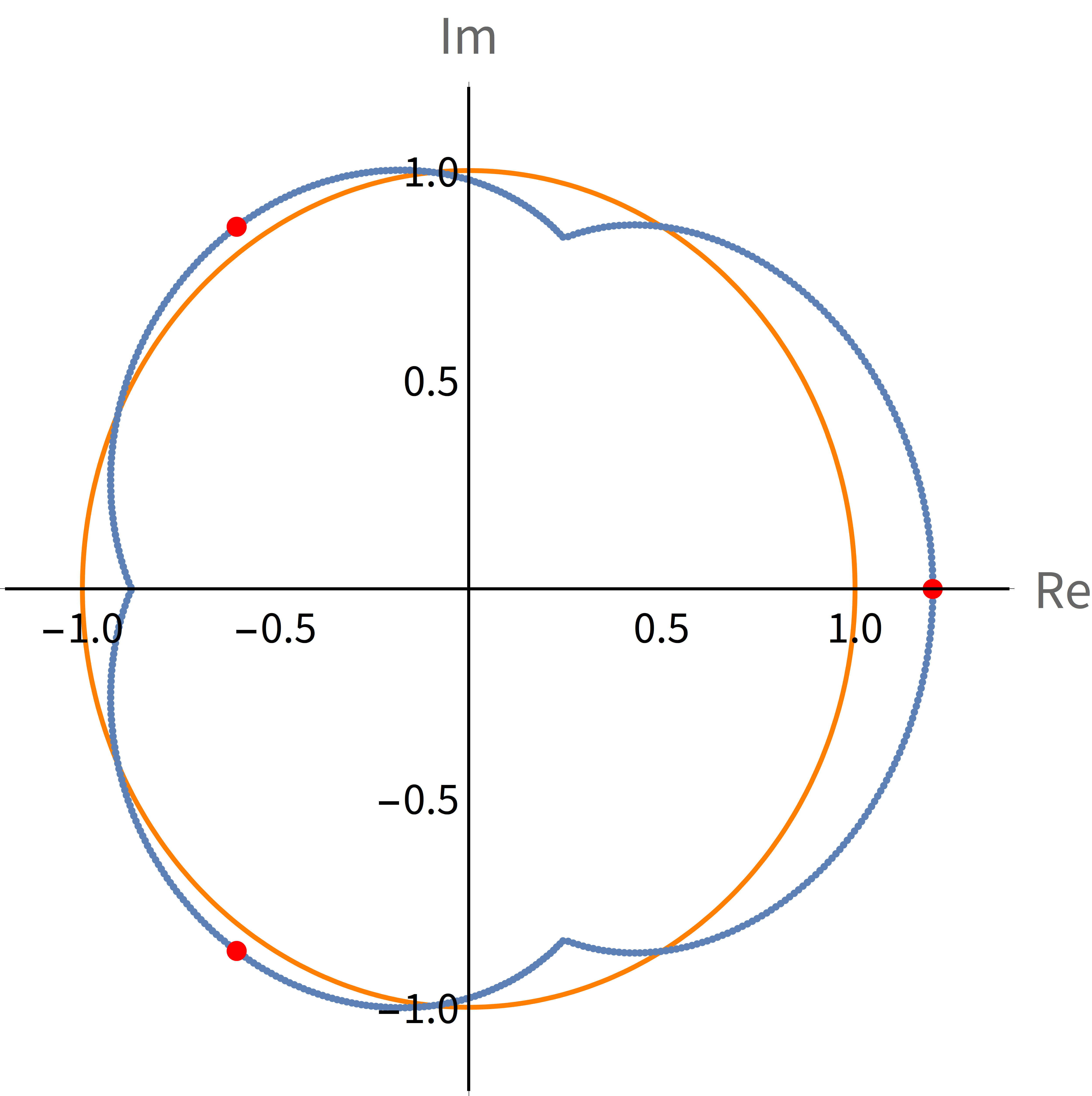}
            \caption{The blue curve is the generalized Brillouin zone($GBZ$) of a $1$-dimensional non-Hermitian HN model with $t_{-1}=3.0$, $t_1=1.0$ and $t_2 = 2.0$. The orange curve is the unit circle. The three red points are three solutions of equation $C(\beta)=0$, and one of them is real. In this non-Hermitian system, $|\beta|$ varies from $|\beta_{min}| < 1$ to $|\beta_{max}|>1$ continuously.} \label{FIG1}
        \end{figure}

        On the other hand, Schr\"{o}dinger equation of a free particle in curved space is given by taking the Laplace operator in curved space
            \begin{equation}
                \frac{\hbar^2}{2M} \left[ \frac{-1}{\sqrt{g}} \partial_i g^{ij} \sqrt{g} \partial_j \right] \phi^{\prime} = E^{\prime} \phi^{\prime} , \label{6}
            \end{equation}
        where $g=det(\mathbf{g})$, $\mathbf{g}$ is the metric and $g^{ij}$ is the $(i,j)$ component of $\mathbf{g}^{-1}$, and $E^{\prime}$ is the eigenenergy of the state. For a $1$-dimensional manifold, $\mathbf{g}=g_{11}=g(y)=g$. Substituting this metric into Eq.\eqref{6}, we get Schr\"{o}dinger equation in $1$-dimensional curved space:
            \begin{equation}
                \left[\frac{1}{g(y)} \partial_y^2 - \frac{\partial_y g(y)}{2g^2(y)} \partial_y \right]\phi^{\prime} = -\frac{2ME^{\prime}}{\hbar^2} \phi^{\prime} . \label{7}
            \end{equation}
        Comparing Eq.\eqref{7} with Eq.\eqref{5}, it is a natural choice to take $g(y)=\frac{1}{B(qy)^2}$. If we take this choice, the coefficient of the first-order derivative term in Eq.\eqref{7} becomes $- \frac{\partial_y g(y)}{2g^2(y)} = Bq^2y$ and the $Cqy$ part of Eq.\eqref{5} is not contained in Eq.\eqref{7}. We can neglect the $Cqy$ term only if $C=0$, but for a general case of $1$-dimensional non-Hermitian HN model, $C \not= 0$. When $C = 0$, $e^{-q}$ has discrete solutions. However, for a general case, according to non-Bloch band theory, $e^{-q}=|\beta|$ \footnote[1]{In non-Bloch band theory, the wavefunction has the form $\psi_n = \beta^n \phi$ with $\beta \in \mathbb{C}$ \cite{17,19}. In this paper, we assume that the wavefunction has the form $\psi_n = e^{-qn} \phi_n$ with $q \in \mathbb{R}$. By comparing them, we get $e^{-q}=|\beta|$} usually takes continuous value from $|\beta|_{min}$ to $|\beta|_{max}$ corresponding to different eigenstates(an example is showed in Fig.\ref{FIG1}) \cite{19}. There is a special case, $|\beta|_{max} = |\beta|_{min} = e^{-q_0} $ and the generalized Brillouin zone~($GBZ$) of this systems is a circle. In this case, the Hamiltonian $H$ given by Eq.\eqref{1} can be transformed to a Hermitian Hamiltonian $\bar{H}$ by a similarity transformation, $\bar{H} = S H S^{-1}$ with a diagonal matrix $S = diag \{e^{q_0} , e^{2q_0}, \cdots , e^{Lq_0}  \}$ and $C = 0$ for all eigenmodes of the non-Hermitian system~(e.g. non-Hermitian HN model with only the nearest hopping) \footnote[2]{If the $GBZ$ of a non-Hermitian system given by Eq.(1) is a circle, all wavefunctions of this system have the same decay factor $e^{- q_0}$. Hence, after taking the similarity transformation, $\bar{H} = S H S^{-1}$, with a diagonal matrix $S = diag \{e^{q_0} , e^{2q_0}, \cdots , e^{Lq_0}  \}$, all wavefunctions of $\bar{H}$ are extended and the $GBZ$ of $\bar{H}$ is the unit circle. According to Ref.\cite{21}, since $\bar{H}$ has no skin effect, the spectrum of $\bar{H}$ under PBC is a line but not a loop in the complex plane~(since $GBZ$ is the unit circle, the spectrum under PBC and OBC are the same if we do not consider the topological state). Since $\bar{H}_{i,j} = t_{i-j} e^{(i-j) q_0}$, the spectrum of $\bar{H}$ is $\bar{E}(k) = \sum_j t_{j} e^{ q_0 j} e^{i k j} $. Since $t_j \geqslant 0$ and $e^{q_0 j} >0$, if $\bar{E}(k)$ gives a line in complex plane when $k$ varies from $0$ to $2 \pi$, $\bar{E}(k) = \bar{E}(-k) \in \mathbb{R}$ for all $k \in [0,2\pi]$. Thus, $t_j e^{q_0 j} = t_{-j} e^{- q_0 j}$, $\bar{H}$ is Hermitian and $C =0$.    }. We call that the non-Hermiticity of such a system, whose Hamiltonian is given by Eq.\eqref{1} and the $GBZ$ is a circle, is erasable. For general cases, $C \not= 0$ and the non-Hermiticity of these systems is non-erasable. So, it is not enough to use a $1$-dimensional manifold to establish the duality. In fact, the coefficients of the first-order derivative term and the second-order derivative term in Eq.\eqref{5} are two independent functions. To construct them, we need at least two independent functions in the metric. We introduce a new coordinate dimension and assume that the metric has the form:
            \begin{equation}
                \mathbf{g}=
                    \begin{pmatrix}
                        g_{yy}(y)& 0& \\
                        0&  g_{zz}(y)&
                    \end{pmatrix},
                    \label{8}
            \end{equation}
        where $g_{yy}$ and $g_{zz}$ are independent of variable $z$. It ensures that Schr\"{o}dinger equation on the $2$-dimensional manifold can be corresponded to a $1$-dimensional Schr\"{o}dinger equation (See Appendix A for details).
        \par

        Substituting the $2$-dimensional metric Eq.\eqref{8} into Eq.\eqref{6}, and comparing with Eq.\eqref{5}, we get functions $g_{yy}(y)$ and $g_{zz}(y)$ analytically (Details are given in  Appendix A) as
            \begin{equation}
                \mathbf{g}=
                   \begin{pmatrix}
                       \frac{1}{B(qy)^2}& 0& \\
                       0&  y^{\frac{2C}{Bq}}
                   \end{pmatrix} .
                   \label{9}
            \end{equation}
        Thus, we get a correspondence. A wavefunction $\psi(x)=e^{-qx} \phi(x)$ of a $1$-dimensional non-Hermitian HN model with eigenenergy $E$ can correspond to a wavefunction $\phi^{\prime}(y)$ of a free particle in a $2$-dimensional curved space, whose metric is given by Eq.\eqref{9}, with eigenenergy $E^{\prime}$. The concrete correspondence relations about wavefunctions and eigenenergies are given in Appendix G. Eq.\eqref{9} shows that the non-erasable non-Hermiticity of the dynamics of free particle in $y$ direction can come from the other dimension since only $g_{zz}$ contains parameter $C$ in metric $\mathbf{g}$.
        \par

        For general cases, the phase of $\phi(x)$ may changes not very smoothly~($\theta$ is finite). However, in the process of establishing the correspondence relation~(Taylor expansion), we request that the the phase of $\phi(x)$ changes smoothly~($|\theta|$ is small). Just a part of eigenstates of the non-Hermitian system satisfy this condition. From Eq.\eqref{9}, we find the metric contains parameters $B$, $C$ and $q$, which do not contain $\theta$ directly. Thus, we assume that the correspondence relation obtained~(the metric, correspondence about wavefunctions and eigenenergies between two systems) can be generalized to the case that $|\theta|$ is finite. Under this assumption, for a wavefunction $\psi(x)=e^{-qx}|\phi(x)|e^{-i \theta x}$ with a finite $|\theta|$ of a $1$-dimensional non-Hermitian HN model, the metric of the dual curved space is still $\mathbf{g}$~(Eq.\eqref{9}) and $\psi(x)$ corresponds to $\phi^{\prime}(y)$, $A$ corresponds to $-\frac{2ME^{\prime}}{\hbar^2}$. In Appendix G, we discuss why this generalization is valid in details and give the specific correspondence relations.

        \par

        It should be noticed that the metric $\mathbf{g}$ depends on parameter $q$. However, different solutions of Schr\"{o}dinger equation about non-Hermitian HN model may have different $q$ (See the example in Fig.\ref{FIG1}). To solve this problem, we introduce a tuning parameter $w$ to represent $q$. We need a $2$-dimensional manifold with a parameter $w$ as dual curved space, whose metric is $\mathbf{g}_1(w)=diag \{ g_{yy},g_{zz} \}$, and $g_{yy}=\frac{1}{B(wy)^2}$, $g_{zz}=y^{\frac{2C}{Bw}}$.  We use $\mathcal{A}_1$ to denote the manifold with metric $\mathbf{g}_1$. There exist two coordinates $y$, $z$ and a parameter $w$ in $\mathcal{A}_1$, and $w \in [q_{min},q_{max}]$. 
        \par

         When we take $q_0$ as the value of parameter $w$, we get a $2$-dimensional manifold with coordinates $y$, $z$ and metric $\mathbf{g}_1(q_0)$. The wavefunction $\psi(x)=e^{-q_0 x} \phi(x)$ with eigenenergy $E$ of the non-Hermitian HN model correspond to a wavefunction $\phi^{\prime}(y)$ of a free particle with eigenenergy $E^{\prime}$ in the $2$-dimensional manifold. $\psi(x)$ corresponds to $\phi^{\prime}(y)$ and $A=E-\sum_j t_j e^{q_0 j}$ corresponds to $-\frac{2 M E^{\prime}}{ \hbar^2}$. Thus, as $w$ changing from $q_{min}$ to $q_{max}$, each solution of Schr\"{o}dinger equation of the $1$-dimensional non-Hermitian HN model with decay factor $e^{-q_0}$ corresponds to a solution of Schr\"{o}dinger equation of a Hermitian Hamiltonian in a $2$-dimensional manifold, which is obtained by taking $q_0$ as the value of parameter $w$, and its metric is given by $\mathbf{g}_1 (q_0)$. Now, we give the complete duality between  non-Hermitian HN model in $1$-dimensional flat space and a Hermitian Hamiltonian in $2$-dimensional manifold $\mathcal{A}_1$ with metric $\mathbf{g}_1$.

        \par
        Now, we discuss the boundary condition of the manifold $\mathcal{A}_1$. The coordinate $y$ comes from the coordinate transformation $y=e^{qx}$. $x \in [1,L]$ is the coordinate of the $1$-dimensional non-Hermitian HN model with open boundary condition(OBC). Thus, as a coordinate of the manifold, $y \in [e^q,e^{qL}]$ and we take OBC in $y$ endpoints. In the proof of Eq.\eqref{9}~(Appendix A), we take the $k_z = 0$ mode to simplify the equation. So, in $z$ direction, periodic boundary condition~(PBC) is taken. Since $w$ is a parameter, it is irrelevant to the dynamics of the free particle system and the Schr\"{o}dinger equation does not contain $\frac{\partial}{\partial w}$ and $\frac{\partial^2}{\partial w^2}$ term. Hence, we do not need to consider the boundary condition about $w$ and we just need $w \in [q_{min}, q_{max}]$.

    \section{High-dimensional case}
        Now, we consider the case of high-dimensional non-Hermitian HN model with open boundary condition. The Hamiltonian can be given by
           \begin{equation}
               H=\sum_{n_1,n_2, \cdots ,n_d} \sum_{j_1,j_2, \cdots ,j_d} t_{j_1,j_2, \cdots ,j_d} c_{n_1+j_1, \cdots , n_d+j_d}^{\dagger} c_{n_1, \cdots ,n_d} ,
               \label{10}
           \end{equation}
        where $n_1,\cdots,n_d$ are indexes of position of the site, $n_i \in [1,L_i]$ for $i=1,2,\cdots,d$~($d\geqslant 2$), and $j_1,\cdots,j_d$ are indexes of hopping length. Similar to $d=1$ case, we assume the wavefunction as
           \begin{equation}
               \psi(x_1, \cdots ,x_d) = e^{-q_1 x_1} e^{-q_2 x_2} \cdots e^{-q_d x_d} \phi(x_1, \cdots ,x_d) ,
               \label{11}
           \end{equation}
        where $e^{-q_i}$ is the decay factor of the $i$th direction, and similar to the $d=1$ case, we consider the eigenstate that $\phi$ changes smoothly with the variation of position here.
        \par

        Then, we introduce a group of new coordinates
           \begin{equation}
               y_1=e^{q_1 x_1}, y_2=e^{q_2 x_2},\cdots,y_d=e^{q_d x_d} ,
               \label{12}
           \end{equation}
        and $y_i \in [e^{q_i},e^{q_i L_i}]$ for $i=1,2,\cdots,d$. Under these coordinate transformations, the Schr\"{o}dinger equation of this system is transformed to (See Appendix B for details)
           \begin{multline}
               \left[\sum_{i=1}^d B_i (q_i y_i)^2 \partial_{y_i}^2  + \sum_{i=1}^d (D_i q_i y_i +B_i q_i^2 y_i ) \partial_{y_i} \right.\\
               \left. + \sum_{m<n} C_{mn} q_m q_n y_m y_n \partial_{y_m} \partial_{y_n} +A \right] \phi =0 ,
               \label{13}
           \end{multline}
        where $B_i>0$ are always held for $i=1,2, \cdots ,d$.
        \par

        In Eq.\eqref{13}, we have $d$ first-order terms, $d$ second-order terms and $\frac{d(d-1)}{2}$ crossing terms. Hence, there are totally $d+\frac{d(d+1)}{2}$ independent functions as their coefficients. However, the metric of a  $d$-dimensional manifold has $\frac{d(d+1)}{2}$ independent functions since the metric must be a symmetric tensor. So, we need to introduce additional $d$ coordinates. In other word, to establish the duality, we need a $2d$-dimensional manifold with its metric. Hence, besides the original coordinates $\{ y_1, \cdots ,y_d\}$, we introduce $d$ new coordinates $\{ z_1, \cdots ,z_d\}$ to construct the $2d$-dimensional metric. After getting the $2d$-dimensional metric, we establish the correspondence about the eigenstate $\psi$. Then, like the $d=1$ case, we can also generalize this correspondence relation to all eigenstates of the non-Hermitian system~(Appendix G). There is no doubt that the $2d$-dimensional metric contains $d$ parameters $q_1, \cdots ,q_d$. Thus, similar to the $1$-dimensional case, we need to introduce other $d$ tuning parameters $\{ w_1, \cdots ,w_d\}$ to represent $\{ q_1,q_2, \cdots, q_d \}$ and $w_i \in [q_{i,min},q_{i,max}]$ for $i = 1,2,\cdots,d$. We use $\mathcal{A}_d$ to denote the $2d$-dimensional manifold with $d$ parameters. And then, an eigenstate of a $d$-dimensional non-Hermitian HN model with decay factors $e^{-q_1}, \cdots ,e^{-q_d}$ can correspond to an eigenstate of a free particle system  in a $2d$-dimensional curved space. This $2d$-dimensional manifold can be obtained by taking $q_i$ as the value of parameter $w_i$ for $i=1,2, \cdots, d$.
        \par

        By using similar method, we give the metric of $\mathcal{A}_d$ analytically~(In Appendix C) as
           \begin{equation}
               \mathbf{g}_d (w_1,w_2, \cdots, w_d) = diag\{ \mathbf{gy}_d, \mathbf{gz}_d  \} ,
               \label{14}
           \end{equation}
        where $\mathbf{gy}_d$ and $\mathbf{gz}_d $ are $d \times d$ matrix, with
           \begin{gather}
            \mathbf{gz}_d = diag\{ g_{z_1 z_1}, g_{z_2 z_2}, \cdots ,g_{z_d z_d} \}  \notag
            \\ = diag\{ y_1^{\frac{A_1}{w_1}}, y_2^{\frac{A_2}{w_2}}, \cdots ,y_d^{\frac{A_d}{w_d}} \} ,
            \label{15}
           \end{gather}
           \begin{multline}
               \mathbf{gy}_d^{-1}=
                  \begin{pmatrix}
                      g^{y_1 y_1} & g^{y_1 y_2} & \cdots & g^{y_1 y_d} \\
                      g^{y_1 y_2} & g^{y_2 y_2} & \cdots & g^{y_2 y_d} \\
                      \vdots      & \vdots      & \ddots & \vdots     \\
                      g^{y_1 y_d} & g^{y_2 y_d} & \cdots        & g^{y_d y_d}
                  \end{pmatrix}
                  \\
                  =
                  \begin{pmatrix}
                       B_1 w_1^2 y_1^2 & \frac{1}{2} C_{12} w_1 w_2 y_1 y_2 & \cdots & \frac{1}{2}  C_{1d} w_1 w_d y_1 y_d \\
                       \frac{1}{2} C_{12} w_1 w_2 y_1 y_2 & B_2 w_2^2 y_2^2 & \cdots & \frac{1}{2}  C_{2d} w_2 w_d y_2 y_d \\
                       \vdots & \vdots & \ddots & \vdots\\
                       \frac{1}{2}  C_{1d} w_1 w_d y_1 y_d & \frac{1}{2}  C_{2d} w_2 w_d y_2 y_d &\cdots &B_d w_d^2 y_d^2
                  \end{pmatrix}.
                \label{17}
           \end{multline}
        $\{ A_1, \cdots ,A_d \}$ are given in Appendix C. If $\{w_1, \cdots ,w_d\}$ are fixed at $\{q_1, \cdots ,q_d\}$, the metric of the $2d$-dimensional manifold is $\mathbf{g}_d (q_1,q_2, \cdots, q_d) $.
        \par

        Now, we establish the duality between $d$-dimensional non-Hermitian HN model in flat space and $2d$-dimensional Hermitian system in curved space completely. For an eigenstate $\psi(x_1, \cdots, x_d)=(\prod_{i=1}^d e^{-q_i x_i}) \phi(x_1, \cdots, x_d)$ of $d$-dimensional HN model with eigenenergy $E$, it corresponds to an eigenstate $\phi^{\prime} (y_1, \cdots, y_d)$ of a free particle in a $2d$-dimensional manifold, which is obtained by taking  $\{ q_1, \cdots, q_d \}$ as the values of parameters $\{ w_1,w_2,\cdots, w_d \}$, and the metric of this manifold is $\mathbf{g}_d (q_1, \cdots, q_d)$.

        \par

         Similar to the $d=1$ case, we take OBC for coordinates $y_1,y_2, \cdots ,y_d$, PBC for coordinates $z_1,z_2, \cdots, z_d$. As for $\{ w_1,w_2, \cdots, w_d \}$, we just need $w_i \in [q_{i,mim},q_{i,max}]$ for $i = 1,2,\cdots,d$.

        \par

         In this section, we study all for the case $d \geqslant 2$. Combining with the results obtained in the previous section, we can find that all of the results (including Eq.\eqref{14} \eqref{15} and \eqref{17}) are valid for $d \geqslant 1$.

    \section{the non-Hermitian system originating from curved space}
       In Sec.$\mathbf{II}$ and Sec.$\mathbf{III}$, we obtain the curved space, which dual to non-Hermitian HN model. On the other hand, if a curved space is given with metric $\mathbf{g}$~(Eq.\eqref{9}) or $\tilde{\mathbf{g}}_d$~(Eq.\eqref{C3} and Eq.\eqref{C18}), we can obtain a non-Hermitian HN model corresponding to it. To get Eq.\eqref{4} and Eq.\eqref{13}, we use Taylor expansion and keep the first and second order terms. Some information about the non-Hermitian system may be lost in this process. This makes the free particle system in a curved space can correspond to more than one non-Hermitian system in flat space. In the following, we show how to obtain a non-Hermitian system from the curved space.

     \subsection{$d=1$ case}
       The metric of the curved space corresponding to the eigenstate of a $1$-dimensional non-Hermitian HN model with decay factor $e^{-q}$ is given by Eq.\eqref{9}, which include given $B$, $C$ and $q$, and where
        \begin{align}
            B=\sum_j \frac{j^2}{2} t_j e^{qj} \label{18} \\
            C=\sum_j -j t_j e^{qj}. \label{19}
        \end{align}
       For simplicity, we only consider two hopping amplitudes $t_1$ and $t_{-1}$. In this case, Eq.\eqref{18} and Eq.\eqref{19} become
        \begin{equation}
            \begin{cases}
                \frac{1}{2} t_{-1} e^{-q} + \frac{1}{2} t_1 e^q = B \\
                t_{-1} e^{-q} - t_1 e^q = C
            \end{cases},
            \label{20}
        \end{equation}
       and their solutions are
        \begin{equation}
            \begin{cases}
                t_{-1}=\frac{2B+C}{2} e^q \\
                t_1=\frac{2B-C}{2} e^{-q} \\
                e^{-q} = \sqrt{\frac{(2B+C)t_1}{(2B-C)t_{-1}}}
            \end{cases}.
            \label{21}
        \end{equation}
         However, for $1$-dimensional non-Hermitian HN model with the nearest hopping, the absolute value of the decay factor given by non-Bloch band theory is \cite{19}
         \begin{equation}
            |\beta| = e^{-q} = \sqrt{\frac{t_1}{t_{-1}}}.
            \label{22}
         \end{equation}
        Comparing Eq.\eqref{21} with Eq.\eqref{22}, we find that Eq.\eqref{21} gives the correct results only if $C=0$. Thus, for a metric $\mathbf{g}$ given by Eq.\eqref{9} with $B>0$, $q \in \mathbb{R}$ and $C=0$, we can construct a $1$-dimensional non-Hermitian HN model with the nearest hopping, which are $t_{-1}=B e^q$ and $t_1=B e^{-q}$, respectively.
        \par

        For $C \not= 0$ cases, the non-Hermitian system constructed must have at least $3$ hopping amplitudes. Here, we choose $t_{-1}$, $t_1$ and $t_2$. Eq.\eqref{18} and Eq.\eqref{19} are written as
         \begin{align}
             \frac{1}{2} t_{-1} e^{-q} + \frac{1}{2} t_1 e^q + 2 t_2 e^{2q} = B
             \label{23}
             \\
             t_{-1} e^{-q} -t_1 e^{q} - 2 t_2 e^{2q} = C .
             \label{24}
         \end{align}
        \par

        According to non-Bloch band theory, the continuous spectrum of this non-Hermitian system is given by $E(\beta)= t_{-1} \beta + t_1 \beta^{-1} + t_2 \beta^{-2}$ with $\beta \in GBZ$~(generalized Brillouin zone). If $\beta \in GBZ$, there must exist a $\theta \in [0,2\pi]$ such that $\beta e^{i \theta} \in GBZ$ and $E(\beta) = E(\beta e^{i \theta})$\cite{19,26}, which can be expressed as
         \begin{equation}
             t_{-1}(1-e^{i \theta}) \beta + t_1 (1-e^{-i \theta}) \beta^{-1} + t_2 (1-e^{-2i \theta}) \beta^{-2} =0 .
             \label{25}
         \end{equation}
        It is a physical request that the non-Hermitian system must have a state with decay factor $|\beta|=e^{-q}$. Since the $GBZ$ of the $1$-dimensional HN model is always symmetric about the real axis, we choose $\theta = \theta_0 \in (0,2\pi)$ and let $\beta = e^{-q} e^{-i \frac{\theta_0}{2}}$. Then, substituting them into Eq.\eqref{25}, we get
         \begin{multline}
            t_{-1}(1-e^{i \theta_0}) e^{-q} e^{-i \frac{\theta_0}{2}} + t_1 (1-e^{-i \theta_0}) e^{q} e^{i \frac{\theta_0}{2}} \\
            + t_2 (1-e^{-2i \theta_0}) e^{2q} e^{i \theta_0} =0.
            \label{26}
         \end{multline}
        Considering Eq.\eqref{23}, Eq.\eqref{24} and Eq.\eqref{26}, we can get the value of $t_{-1}$, $t_1$ and $t_2$.
        \par

        Through the non-Bloch band theory, we can obtain the $GBZ$ of the $1$-dimensional non-Hermitian HN model with hopping amplitudes $t_{-1}$, $t_1$ and $t_2$. The next step is checking whether $\beta =e^{-q} e^{-i \frac{\theta_0}{2}}$ belongs to this $GBZ$, because Eq.\eqref{26} can not guarantee that $\beta \in GBZ$. If $\beta \in GBZ$, this $1$-dimensional non-Hermitian system with hopping amplitudes $t_{-1}$, $t_1$ and $t_2$ is the system corresponding to the curved space. If $\beta = e^{-q} e^{-i \frac{\theta_0}{2}}$ does not belong to the $GBZ$ obtained, we choose $\theta = \theta_1 \neq \theta_0$ and then solve new values of $t_{-1}$, $t_1$ ,$t_2$ and get a new $GBZ$. We repeat the preceding process until the $\beta$ chosen belongs to the new $GBZ$ of the non-Hermitian system.

      \subsection{An example of $d=1$ case}
         We take an example to elucidate how to find the $1$-dimensional non-Hermitian HN model from a given curved space.
        \par

        Consider a curved space with coordinates $y$ and $z$, whose metric is given by Eq.\eqref{9} with $B=4$, $C=-2$ and $e^q =2$. Substituting these parameters into Eq.\eqref{23} and Eq.\eqref{24}, we get
         \begin{align}
             \frac{1}{4} t_{-1} + t_1 + 8 t_2 = 4  \label{27}
             \\
             \frac{1}{2} t_{-1} - 2t_1 - 8 t_2 = -2 \label{28} .
         \end{align}
        We take $\theta_0 = \pi$, and $\beta = e^{-q} e^{-i \frac{\theta_0}{2}} = -\frac{i}{2}$, Eq.\eqref{26} is simplified as
         \begin{equation}
             t_{-1} - 4 t_1 = 0 .
             \label{29}
         \end{equation}
        Solving Eq.\eqref{27}, Eq.\eqref{28} and Eq.\eqref{29}, we get
         \begin{equation}
             t_{-1}=4 \qquad t_1=1 \qquad t_2=\frac{1}{4} .
             \label{30}
         \end{equation}
        \par

\begin{figure}
            \centering
            
            \subfigure[]{
                \includegraphics[scale=0.51]{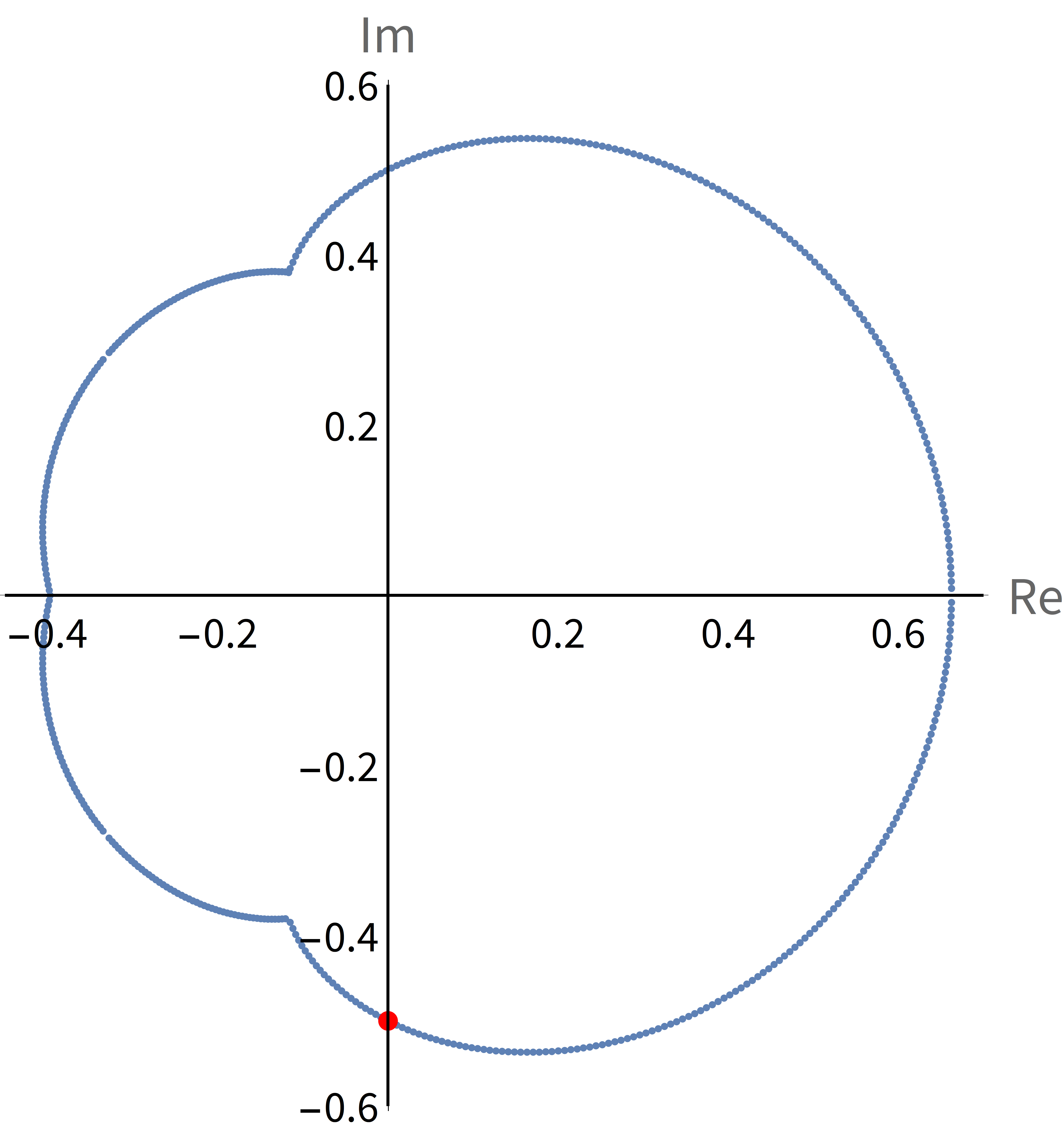} \label{fig2a}}
                \qquad
            \subfigure[]{
                \includegraphics[scale=0.5]{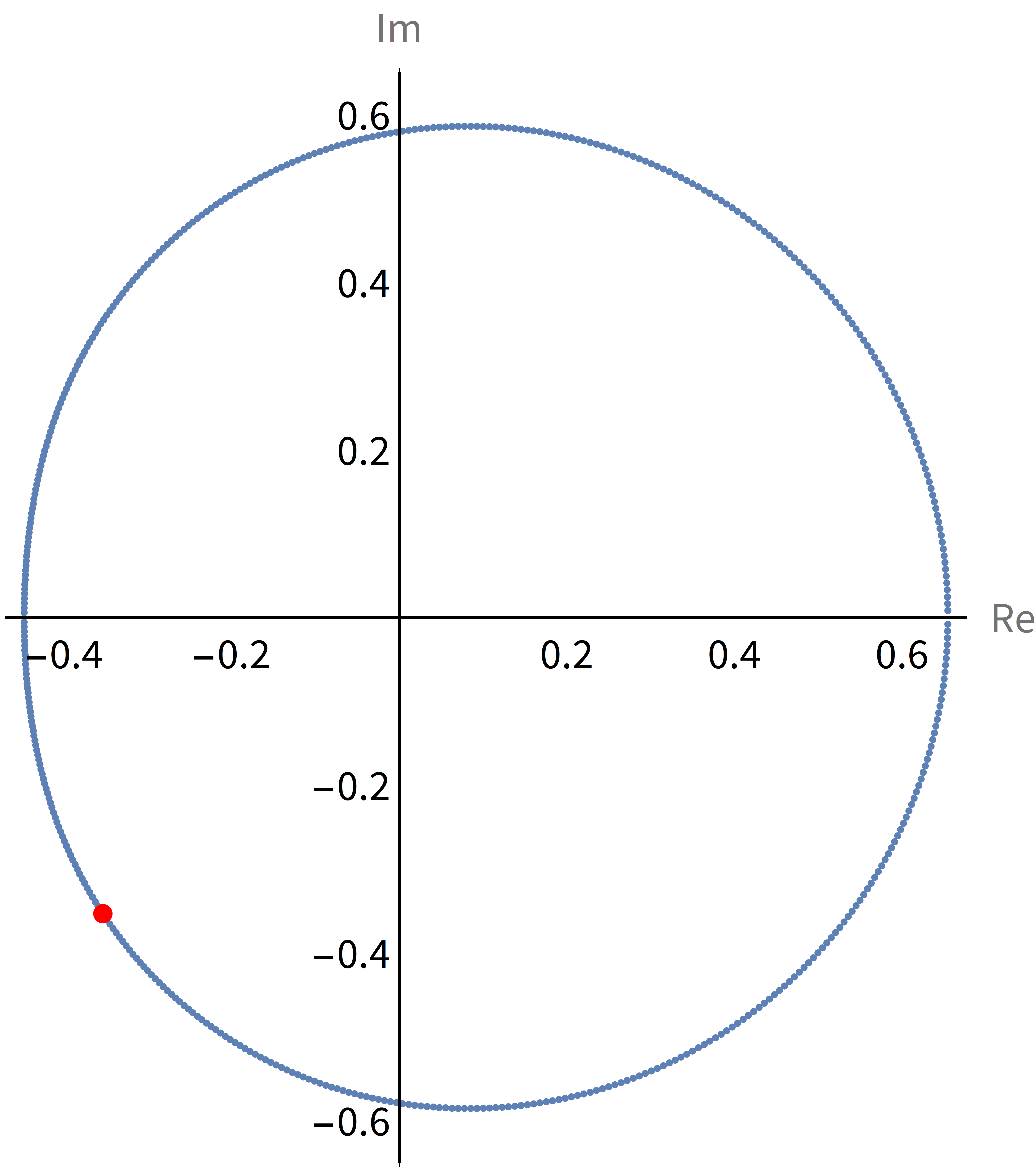} \label{fig2b}}
            
            \caption{(a).The blue curve in the complex plane is the $GBZ$ of the $1$-dimensional non-Hermitian HN model with $t_{-1}=4$, $t_1=1$ and $t_2=\frac{1}{4}$. The red point is $\beta=-\frac{i}{2}$, which belongs to this $GBZ$. (b).The blue curve is the $GBZ$ of the $1$-dimensional non-Hermitian HN model whose hopping amplitudes are given by Eq.\eqref{s} The red point is $\beta=-\frac{\sqrt{2}}{4} - i \frac{\sqrt{2}}{4}$, which belongs to the $GBZ$ of corresponding non-Hermitian HN model}\label{FIG2}
         \end{figure}

        The $GBZ$ of the $1$-dimensional non-Hermitian HN model with hopping amplitude $t_{-1}=4$, $t_1=1$ and $t_2=\frac{1}{4}$ is showed in Fig.\ref{fig2a}. And $\beta=-\frac{i}{2}$ belongs to this $GBZ$~(See Fig.\ref{fig2a}). Thus, this $1$-dimensional non-Hermitian HN model~(with $t_{-1}=4$, $t_1=1$ and $t_2=\frac{1}{4}$) has been constructed from the given curved space~(with $B=4$, $C=-2$ and $e^q =2$). 
        \par


        \par

        In fact, the non-Hermitian system corresponding to this curved space is not unique. For this case, we can also choose $\theta_0 = \frac{3 \pi}{2}$ and the hopping amplitudes can be solved as 
          \begin{equation}
              t_{-1}=\frac{2 (4+3 \sqrt{2})}{2+ \sqrt{2}}   \qquad  
              t_1=\frac{4+ 5 \sqrt{2}}{2 (2+ \sqrt{2})}   \qquad
              t_2= \frac{1}{2 (2+ \sqrt{2})}.
              \label{s}
          \end{equation}
        And we can find $\beta= e^{-q} e^{-i \frac{\theta_0}{2}} = -\frac{\sqrt{2}}{4} - i \frac{\sqrt{2}}{4}$ belongs to the $GBZ$ of the $1$-dimensional non-Hermitian HN model with these hopping amplitudes~(Fig.\ref{fig2b}).

      \subsection{$d \geqslant 2 $ cases}
        Consider a curved space with  metric $\tilde{\mathbf{g}}_d = diag\{ \mathbf{gy}_d, \mathbf{gz}_d \}$ given in Appendix C, and a non-Hermitian system constructed from this curved space. It is necessary to request this non-Hermitian system contains an eigenstate with decay factors $e^{-q_1}, e^{-q_2}, \cdots, e^{-q_d}$. In these cases, we need some other relations between the hopping amplitudes in the non-Hermitian Hamiltonian and the decay factors. For $d=1$ case, Non-Bloch band theory supplies this relation. However, for $d \geqslant 2$ cases, non-Bloch band theory is invalid. In general, we cannot construct the $d$-dimensional non-Hermitian HN model from the curved space.
        \par

        For some special cases, we can give the corresponding non-Hermitian HN model. Consider the case $C_{mn} = 0$  for arbitrary $m$ and $n$. The metric of the curved space can be reduced as
         \begin{multline}
             \tilde{\mathbf{g}}_d = diag\{ \frac{1}{B_1 (q_1 y_1)^2}, \frac{1}{B_2 (q_2 y_2)^2}, \cdots, \frac{1}{B_d (q_d y_d)^2}, \\
             y_1^{\frac{2D_1}{B_1 q_1}}, y_2^{\frac{2D_2}{B_2 q_2}}, \cdots, y_d^{\frac{2D_d}{B_d q_d}}   \}.
             \label{31}
         \end{multline}
        According to the definition of $C_{mn}$~(Appendix B), if $C_{mn} = 0$ for all $m$ and $n$, $t_{j_1,j_2, \cdots, j_d}=0$ except $ \{t_{j_1,0,0,\cdots,0}, t_{0,j_2,0,\cdots,0}, \cdots, t_{0,0,\cdots, j_d}\}$. In this case, the Hamiltonian of the $d$-dimensional non-Hermitian HN model can be simplified as
         \begin{equation}
            H=\sum_{n_1,n_2, \cdots, n_d} \sum_{p=1}^d \sum_{j_p} t^{(p)}_{j_p} c_{x_1,x_2, \cdots, x_p+j_p, \cdots, x_d}^{\dagger} c_{x_1,x_2, \cdots, x_d}.
            \label{32}
         \end{equation}
        Fortunately, non-Bloch band theory is valid for these special cases and gives the relation between the hopping amplitudes and decay factors. As the non-Bloch band theory for $1$-dimensional case, the value of decay factor $e^{-q_i}$ is restricted by $GBZ_i$. $GBZ_i$ can be obtained by hopping amplitude $t_{j_i}^{(i)}$ for $i=1,2, \cdots, d$ respectively~(Details are given in Appendix E). Thus, when the parameters $B_i$, $D_i$ and $q_i$ of the curved space are given, the hopping amplitudes $t_{-1}^{(i)}$, $t_{1}^{(i)}$ and $t_{2}^{(i)}$ of the non-Hermitian system can be obtained as $d=1$ case. Repeating this process for $i=1,2,\cdots,d$, we will obtain all hopping amplitudes in non-Hermitian Hamiltonian Eq.\eqref{32}. Finally, the non-Hermitian system corresponding to the curved space is obtained.

    \section{ some special cases}
        \subsection{$q_i=0$ cases}
           All results above is based on $q_i \not= 0$ for all $i=1, \cdots ,d$. If $q_i=0$ for any one $i$, the transformation of coordinate $y_i=e^{q_i x_i}$ becomes invalid. To solve this problem, we take $y_i=x_i$ directly. Under this transformation, $g^{y_i y_i}$ and $g^{y_i y_j} (j\not=i)$ become $B_i$ and $\frac{1}{2} C_{ij} w_j y_j$ respectively. So, all factors $w_i y_i$ are set to 1 in $\mathbf{gy}_d^{-1}$. And $g_{z_i z_i}$ becomes $e^{A_i y_i}$ in $\mathbf{gz}_d$~(See Appendix D for details).
           \par

           We emphasize that $q_i=0$ does not mean that the manifold $\mathcal{A}_d$ is flat in the $i$th direction, because the off-diagonal terms about coordinate $y_i$ of $\mathbf{gy}_d$ may not be $0$. If $\mathcal{A}_d$ is flat in the $i$th direction, $g^{ij}=0$ are held for all $j\not=i$, otherwise $q_i=0$ only means the wavefunction is extended in the $i$th direction.

        \subsection{The Hermitian limit}
           For Hermitian $d$-dimensional HN model, the corresponding manifold $\mathcal{A}_d$ degenerates into a flat manifold.  In this case, $\mathbf{gy}_d=diag\{ \frac{1}{B_1}, \cdots ,\frac{1}{B_d} \}$ and $\mathbf{gz}_d=I_{d \times d}$ (Appendix D). The metric $\mathbf{g}_d=diag\{ \mathbf{gy}_d,\mathbf{gz}_d \}$ is a diagonal matrix and $\mathcal{A}_d$ is a flat manifold.

        \subsection{System with spatial symmetry}
           For a general case, $q_1, \cdots ,q_d$ are $d$ independent parameters and we introduce $w_1, \cdots ,w_d$ to represent them. However, if the system has spatial symmetry, there will be some restrictions on these parameters. If the symmetry group $G$ gives $l$($l < d$) restrictions on $\{ q_1, \cdots ,q_d\}$, then, only $d-l$ parameters in $\{ q_1, \cdots ,q_d\}$ are independent. Thus, we only need to introduce $\{ w_1, \cdots ,w_{d-l} \}$ to represent them. In this case, $\mathcal{A}_d$ just need to contain $d-l$ parameters.
           \par

           For example, the  wavefunction of $2$-dimensional non-Hermitian HN model has two decay factors, $e^{-q_1}$ and $e^{-q_2}$. If this system has mirror symmetry $\mathcal{M}$ about the line $x_1=x_2$, the wavefunction of the eigenstate has the property $\psi(x_1,x_2)=\psi(x_2,x_1)$. Since $\psi(x_1,x_2)=e^{-q_1 x_1}e^{-q_2 x_2} \phi(x_1,x_2)$ and $\psi(x_1,x_2)=\psi(x_2,x_1)$ are satisfied, the two parameters $q_1$ and $q_2$ must satisfy $q_1 = q_2$. Thus, we only need one tuning parameter $w$ to represent them. In this case,  $\mathcal{A}_2|_{\mathcal{M}}$ is a $4$-dimensional manifold with coordinates $\{ y_1, y_2, z_1, z_2\}$ and parameter $w$.

        \subsection{A Defective Hamiltonian}
           If the Hamiltonian of non-Hermitian system is defective, eigenstates coalesce and eigenenergy degenerate. For a simplest case, we can find the metric of the curved space corresponding to the non-Hermitian system.
           \par

           Consider the $1$-dimensional non-Hermitian system with only on-site energy $t_0$ and the left nearest hopping $t_{-1}$:
             \begin{equation}
                 H= \sum_n t_0 c_n^{\dagger} c_n + t_{-1} c_{n-1}^{\dagger} c_n.
             \end{equation}
            The matrix form of this Hamiltonian is
              \begin{equation}
                  H= \begin{pmatrix}
                      t_0 & t_{-1} \\
                          &  t_0   & t_{-1}  \\
                          &        & \ddots  & \ddots \\
                          &        &         & t_0    & t_{-1} \\
                          &        &         &        & t_0  \\
                  \end{pmatrix}_{L \times L}    ,
              \end{equation}
            where $L$ is the length of this system. Apparently this Hamiltonian is defective with only one coalescent eigenstate $\Psi=(1,0, \cdots, 0)^T$ and eigenenergy $E= t_0$.
            \par

            According to non-Bloch band theory, the decay factor of this eigenstate is\cite{19} $e^{-q} = \sqrt{\frac{t_1}{t_{-1}}} = 0$. Hence, $q \to \infty$. Since $q$ cannot be infinite in a real system, we treat $q$ as a large number when dealing with the system and then take the limit $q \to \infty$. Considering Eq.\eqref{9},
              \begin{equation}
                  \mathbf{g}|_{q \to \infty} = diag \{ 0,1 \}.
              \end{equation}
             The metric of the $2$-dimensional submanifold corresponding to this coalescent eigenstate is degenerate.

     \section{Conclusion and discussion}
        In this article, we give a duality between $d$-dimensional non-Hermitian HN model in flat space and Hermitian system in $2d$-dimensional curved space, and get the metric of the curved space analytically. All results are valid for all $d\geqslant 1$ cases.  Our results present a new origin of non-Hermiticity of the system, and realize a new path to explore high-dimensional non-Hermitian systems in the curved space. We can use an isolated system in curved space to simulate a non-Hermitian system in flat space and realize skin effect or exceptional point effectively. For a given non-Hermitian HN model with skin effect or exceptional point, we can find the metric corresponding to localized bulk states or exceptional state(corresponding to the exceptional point) and construct the curved space, then we can research skin effect or exceptional point by the corresponding Hermitian system in this curved space. On the other hand, we can utilize a low-dimensional non-Hermitian system in flat space to realize and study a high-dimensional system in curved space.
        \par

        In the future, the duality about multi-orbital non-Hermitian systems will be considered. It will be helpful to find a systematic method to research high-dimensional non-Hermitian systems. Furthermore, this duality gives a new way to construct the many body states in non-Hermitian system. The non-Hermitian many body states can be constructed as the states corresponding to the many body states of a Hermitian system in a curved space.
        \par

    \section{acknowledgements}
       The authors thank useful discussion with Yongxu Fu, Haoshu Li and Jihan Hu. This work was supported by
       NSFC Grant No. 11275180.

  \appendix

  \section{$2$-Dimensional Metric $\mathbf{g}$}
  In this appendix, we show how to get $2$-dimensional metric $\mathbf{g}$, Eq.(9).
  \par

  We denote $g=det(\mathbf{g})$, $g^{yy}=\frac{1}{g_{yy}}$, and $g^{zz}=\frac{1}{g_{zz}}$. Eq.(6) is expressed as
      \begin{multline}
          \frac{\hbar^2}{2M} \left[\frac{-1}{\sqrt{g_{yy}g_{zz}}} (\partial_y \frac{\sqrt{g_{yy}g_{zz}}}{g_{yy}} \partial_y + \partial_z \frac{\sqrt{g_{yy}g_{zz}}}{g_{zz}} \partial_z) \right] \phi^{\prime}(y,z) \\
           =E^{\prime} \phi^{\prime}(y,z) .
          \tag{A1}
          \label{A1}
      \end{multline}
  Since functions $g_{yy}$ and $g_{zz}$ are independent of variable $z$, Eq.\eqref{A1} can be simplified as
      \begin{multline}
          \left[\frac{1}{g_{yy}} \partial_y^2 + \frac{1}{g_{zz}} \partial_z^2 + \frac{1}{2\sqrt{g_{yy}g_{zz}}} \sqrt{\frac{g_{yy}}{g_{zz}}} \frac{g_{yy} \partial_y g_{zz} - g_{zz} \partial_y g_{yy}}{g_{yy}^2} \partial_y \right. \\
           \left. + \frac{2ME^{\prime}}{\hbar^2} \right] \phi^{\prime}(y,z) = 0 .
          \tag{A2}
          \label{A2}
      \end{multline}
  From Eq.\eqref{A2}, we have $\phi^{\prime}(y,z)=\phi^{\prime}(y)e^{ik_z z}$. For simplicity, considering the $k_z=0$ mode, we obtain an equation about $\phi(y)$:
      \begin{multline}
          \left[\frac{1}{g_{yy}} \partial_y^2 + \frac{1}{2\sqrt{g_{yy}g_{zz}}} \sqrt{\frac{g_{yy}}{g_{zz}}} \frac{g_{yy} \partial_y g_{zz} - g_{zz} \partial_y g_{yy}}{g_{yy}^2} \partial_y  \right. \\
           \left. + \frac{2ME^{\prime}}{\hbar^2} \right] \phi^{\prime}(y) = 0 .
          \tag{A3}
          \label{A3}
      \end{multline}
  To establish the duality, we let Eq.\eqref{A3} equal Eq.(5). Comparing the coefficients of the second-order terms of Eq.\eqref{A3} and Eq.(5), we get
      \begin{equation}
          g_{yy}=\frac{1}{B(qy)^2} .
          \tag{A4}
          \label{A4}
      \end{equation}
  Then, substituting $g_{yy}$ into the coefficient of the first-order term in Eq.\eqref{A3}, we have
      \begin{equation}
          \frac{-1}{2\sqrt{g_{yy}g_{zz}}} \sqrt{\frac{g_{yy}}{g_{zz}}} \frac{g_{zz} \partial_y g_{yy}}{g_{yy}^2}= \frac{-\partial_y g_{yy}}{2g_{yy}^2}=Bq^2y .
          \tag{A5}
          \label{A5}
      \end{equation}
  Comparing Eq.\eqref{A3} with Eq.(5), we get
      \begin{equation}
          \frac{1}{2\sqrt{g_{yy}g_{zz}}} \sqrt{\frac{g_{yy}}{g_{zz}}} \frac{g_{yy} \partial_y g_{zz}}{g_{yy}^2} =B(qy)^2\frac{\partial_y g_{zz}}{2g_{zz}} = Cqy .
          \tag{A6}
          \label{A6}
      \end{equation}
  From Eq.\eqref{A6}, we have
      \begin{equation}
          g_{zz}=cy^{\frac{2C}{Bq}} ,
          \tag{A7}
          \label{A7}
      \end{equation}
  where $c > 0$ is constant. We set $c=1$ and take $g_{zz}=y^{\frac{2C}{Bq}}$.
  \par

  From above, the $2$-dimensional metric, Eq.(9), is obtained.

\section{Proof of Eq.(13)}

 In lattice model, we assume that $\Psi=\{ \psi_{n_1,n_2, \cdots ,n_d} \}$ is the wavefunction of Hamiltonian Eq.(10) with eigenenergy $E$. The Schr\"{o}dinger equation $H \Psi = E \Psi$ can be written as
    \begin{equation}
        \sum_{j_1, \cdots ,j_d} t_{j_1, \cdots ,j_d} \psi_{n_1-j_1, \cdots ,n_d-j_d}  = E \psi_{n_1, \cdots ,n_d} .
        \tag{B1}
        \label{B1}
    \end{equation}
    In general, the wavefunctions take form as
     \begin{equation}
         \psi_{n_1, \cdots ,n_d}=(\prod_{i=1}^d e^{-q_i n_i}) \phi_{n_1,\cdots ,n_d}
         \tag{B2}
         \label{B2}
     \end{equation}

  The extended part of the wavefunction, $\phi$, satisfy
     \begin{equation}
         \sum_{j_1, \cdots ,j_d} t_{j_1, \cdots ,j_d} \phi_{n_1-j_1, \cdots ,n_d-j_d} e^{q_1 j_1} \cdots e^{q_d j_d}=E \phi_{n_1,\cdots ,n_d} .
         \tag{B3}
         \label{B3}
     \end{equation}
      Similar to the $1$-dimensional case, we make the position variables continue and take $\phi_{n_1, \cdots ,n_d}=\phi(x_1, \cdots ,x_d)|_{\{x_1, \cdots ,x_d\}=\{n_1, \cdots ,n_d\}}$ (set lattice constant be $1$). Then, Eq.\eqref{B3}  becomes
     \begin{multline}
      \sum_{j_1, \cdots ,j_d} t_{j_1, \cdots ,j_d} \phi(x_1-j_1, \cdots ,x_d-j_d) e^{q_1 j_1} \cdots e^{q_d j_d} \\
      =E \phi(x_1, \cdots ,x_d) .
      \tag{B4}
      \label{B4}
     \end{multline}
  Expanding $\phi(x_1-j_1, \cdots ,x_d-j_d)$ to the second order at $(x_1, \cdots ,x_d)$, Eq.\eqref{B4} becomes
     \begin{multline}
         \sum_{j_1, \cdots ,j_d} t_{j_1, \cdots ,j_d} (\prod_{i=1}^d e^{q_i j_i}) \left[\phi- \sum_{m=1}^d j_m \partial_{x_m}\phi  \right. \\
          +\sum_{m=1}^d \frac{j_m^2}{2}\partial_{x_m}^2 \phi
          \left.  + \sum_{m < n} j_m j_n \partial_{x_m} \partial_{x_n} \phi \right]= E \phi .
         \tag{B5}
         \label{B5}
     \end{multline}
  Similar to the $d=1$ case, we still keep Taylor expansion at the quadratic terms here. Because it can make the solution of Eq.\eqref{B5} return to plane wave naturally and continuously in the Hermitian limit.
  \par

  Define:
     \begin{equation}
         B_m= \sum_{j_1, \cdots ,j_d} \frac{j_m^2}{2}t_{j_1, \cdots ,j_d} (\prod_{i=1}^d e^{q_i j_i}) ,
         \tag{B6}
         \label{B6}
     \end{equation}
     \begin{equation}
         D_m=\sum_{j_1, \cdots ,j_d} -j_m t_{j_1, \cdots ,j_d} (\prod_{i=1}^d e^{q_i j_i}) ,
         \tag{B7}
         \label{B7}
     \end{equation}
     \begin{equation}
         C_{mn}=\sum_{j_1, \cdots ,j_d} j_m j_n t_{j_1, \cdots ,j_d} (\prod_{i=1}^d e^{q_i j_i}) ,
         \tag{B8}
         \label{B8}
     \end{equation}
     \begin{equation}
         A=\sum_{j_1, \cdots ,j_d} t_{j_1, \cdots ,j_d} (\prod_{i=1}^d e^{q_i j_i})-E .
         \tag{B9}
         \label{B9}
     \end{equation}
  Because magnetic field is absent here, $t_{j_1, \cdots ,j_d}\geqslant 0$. Thus $B_m>0$ for $m=1,2, \cdots ,d$. Then, Eq.\eqref{B5} is simplified as
    \begin{equation}
         \left[\sum_{i=1}^d B_i \partial_{x_i}^2 + \sum_{i=1}^d D_i \partial_{x_i} + \sum_{m<n} C_{mn} \partial_{x_m} \partial_{x_n} +A \right]\phi = 0 .
         \tag{B10}
         \label{B10}
     \end{equation}
  Taking the coordinates transformation Eq.(12), we have $\frac{\partial \phi}{\partial x_i}= q_i y_i \frac{\partial \phi}{\partial y_i}$ and $\frac{\partial^2 \phi}{\partial x_i^2}= (q_i y_i)^2 \frac{\partial^2 \phi}{\partial y_i^2}+q_i^2 y_i\frac{\partial \phi}{\partial y_i} $ for a given function $\phi$ about variables ${x_1,x_2,\cdots,x_d}$. Under this transformation, Eq.(13) is obtained from Eq.\eqref{B10}.

\section{Metric $\mathbf{g}_d$}
 In this appendix, we show how to obtain the metric $\mathbf{g}_d$ analytically,
 \par

  From the main text, we know the metric $\mathbf{g}_d$ contains $2d$ coordinates $\{ y_1, \cdots, y_d, z_1, \cdots, z_d \}$ and $d$ parameters $\{ w_1, \cdots, w_d \}$. To make sure that the Schr\"{o}dinger equation in the $2d$-dimensional curved space, which is obtained by taking $\{ q_1, \cdots, q_d \}$ as the values of $\{ w_1,w_2,\cdots, w_d \}$, can be corresponded to a Schr\"{o}dinger equation of $d$-dimensional HN model, there must exist no crossing derivative term~($\partial_{y_i} \partial_{z_j}$ for arbitrary $i$, $j$) in the $2d$-dimensional Schr\"{o}dinger equation. Hence, $\mathbf{g}_d$ is a block diagonal matrix, i.e.
    \begin{equation}
      \mathbf{g}_d (w_1, w_2, \cdots, w_d) = diag\{ \mathbf{gy}_d, \mathbf{gz}_d \} .
      \notag
    \end{equation}
   Then, define $\tilde{\mathbf{g}}_d= \mathbf{g}_d (q_1, q_2, \cdots, q_d)$. We only need to solve $\tilde{\mathbf{g}}_d$ analytically.
  \par

  To ensure that variables $\{y_1, \cdots ,y_d\}$ and $\{z_1, \cdots ,z_d\}$ are separable in the $2d$-dimensional Schr\"{o}ding equation, all terms of $\tilde{\mathbf{g}}_d$ must be independent of variables $z_1, \cdots ,z_d$, and
     \begin{equation}
         \mathbf{gz}_d=diag\{ g_{z_1 z_1}, \cdots ,g_{z_d z_d} \}=diag\{ \frac{1}{g^{z_1 z_1}}, \cdots ,\frac{1}{g^{z_d z_d}} \} .
         \notag
     \end{equation}
     The general form of $\mathbf{gy}_d$ is
     \begin{equation}
         \mathbf{gy}_d=
            \begin{pmatrix}
              g_{y_1 y_1} & g_{y_1 y_2} & \cdots & g_{y_1 y_d} \\
              g_{y_1 y_2} & g_{y_2 y_2} & \cdots & g_{y_2 y_d} \\
              \vdots      & \vdots      & \ddots & \vdots     \\
              g_{y_1 y_d} & g_{y_2 y_d} & \cdots        & g_{y_d y_d}
            \end{pmatrix} ,
            \notag
     \end{equation}
     or
     \begin{equation}
      \mathbf{gy}_d^{-1}=
         \begin{pmatrix}
           g^{y_1 y_1} & g^{y_1 y_2} & \cdots & g^{y_1 y_d} \\
           g^{y_1 y_2} & g^{y_2 y_2} & \cdots & g^{y_2 y_d} \\
           \vdots      & \vdots      & \ddots & \vdots     \\
           g^{y_1 y_d} & g^{y_2 y_d} & \cdots   & g^{y_d y_d}
         \end{pmatrix} .
         \notag
      \end{equation}
  Denoting that $g=det(\tilde{\mathbf{g}}_d)$, and substituting $\tilde{\mathbf{g}}_d$ into Eq.(6), we get
      \begin{multline}
          \frac{\hbar^2}{2M} \left[-\sum_{i=1}^d g^{z_i z_i} \partial_{z_i}^2 - \sum_{i=1}^d g^{y_i y_i} \partial_{y_i}^2 - \sum_{m \not= n} g^{y_m y_n}\partial_{y_m} \partial_{y_n}  \right. \\
           \left.  -\frac{1}{\sqrt{g}} \sum_{n=1}^d (\sum_{m=1}^d \partial_{y_m} (g^{y_m y_n}) \sqrt{g}) \partial_{y_n} \right] \phi^{\prime} =E^{\prime} \phi^{\prime} .
          \tag{C1}
          \label{C1}
      \end{multline}
  Apparently, Eq.\eqref{C1} is variable separable, and $z$-dependent terms all are second-order terms. Thus, it is a natural choice to assume that $\phi^{\prime}(y_1, \cdots ,y_d,z_1, \cdots ,z_d)=e^{i k_1 z_1}e^{i k_2 z_2} \cdots e^{i k_d z_d} \phi^{\prime}(y_1, \cdots ,y_d)$. Considering the $k_1=k_2= \cdots =k_d=0$ mode, Eq.\eqref{C1} is reduced to an equation about $\phi^{\prime}(y_1, \cdots ,y_d)$
      \begin{multline}
          \left[\sum_{i=1}^d g^{y_i y_i} \partial_{y_i}^2 + \sum_{m\not= n} g^{y_m y_n} \partial_{y_m} \partial_{y_n} \right. \\
           \left. + \sum_{n=1}^d \frac{1}{\sqrt{g}} (\sum_{m=1}^d \partial_{y_m} (g^{y_m y_n} \sqrt{g}))\partial_{y_n} + \frac{2ME^{\prime}}{\hbar^2} \right]\phi^{\prime}=0 .
          \tag{C2}
          \label{C2}
      \end{multline}
  By comparing the coefficients of second-order terms and crossing terms in Eq.(13) with those in Eq.\eqref{C2}, we get
      \begin{equation}
          g^{y_i,y_i}=B_i q_i^2 y_i^2
          \notag
      \end{equation}
  and for $m<n$
      \begin{equation}
          g^{y_m y_n}=g^{y_n y_m} =\frac{1}{2} C_{mn} q_m q_n y_m y_n .
          \notag
      \end{equation}
  Thus, $\mathbf{gy}_d^{-1}$ is obtained analytically:
     \begin{multline}
         \mathbf{gy}_d^{-1}= \\
             \begin{pmatrix}
              B_1 q_1^2 y_1^2 & \frac{1}{2} C_{12} q_1 q_2 y_1 y_2 & \cdots & \frac{1}{2}  C_{1d} q_1 q_d y_1 y_d \\
              \frac{1}{2} C_{12} q_1 q_2 y_1 y_2 & B_2 q_2^2 y_2^2 & \cdots & \frac{1}{2}  C_{2d} q_2q_d y_2 y_d \\
              \vdots & \vdots & \ddots & \vdots\\
              \frac{1}{2}  C_{1d} q_1 q_d y_1 y_d & \frac{1}{2}  C_{2d} q_2q_d y_2 y_d & \cdots &B_d q_d^2 y_d^2
             \end{pmatrix} .
          \tag{C3}
          \label{C3}
     \end{multline}
  For a general case, replacing $\{q_1, \cdots ,q_d\}$ by $\{ w_1, \cdots ,w_d \}$, and we obtain Eq.(16) from Eq.\eqref{C3}.
  \par

  Next, we define
     \begin{equation}
         X=
           \begin{pmatrix}
               B_1 & \frac{1}{2} C_{12} & \cdots & \frac{1}{2} C_{1d} \\
               \frac{1}{2} C_{12} & B_2 & \cdots  &\frac{1}{2} C_{2d} \\
               \vdots  &  \vdots  & \ddots  &  \vdots                 \\
               \frac{1}{2} C_{1d} & \frac{1}{2} C_{2d} &  \cdots  &  B_d
           \end{pmatrix} ,
           \tag{C4}
           \label{C4}
     \end{equation}
     \begin{equation}
         c=det(X) ,  \tag{C5} \label{C5}
     \end{equation}
     \begin{equation}
         f(y_1, \cdots ,y_d)= \prod_{i=1}^d g_{z_i z_i}  .
         \tag{C6}
         \label{C6}
     \end{equation}
 It is easy to obtain the determinant of $\mathbf{gy}_d^{-1}$:
     \begin{equation}
         det(\mathbf{gy}_d^{-1})=c\prod_{i=1}^d (q_i y_i)^2 .
         \tag{C7}
         \label{C7}
     \end{equation}
  Thus,
     \begin{equation}
         g=det(\tilde{\mathbf{g}}_d)=det(\mathbf{gy}_d) det(\mathbf{gz}_d)=\frac{f}{c} \prod_{i=1}^d (q_i y_i)^{-2} .
         \tag{C8}
         \label{C8}
     \end{equation}
  Then, we get
     \begin{gather}
         \sqrt{g}=\frac{\sqrt{f}}{\sqrt{c}} \frac{1}{\prod_{i=1}^d q_i y_i}  , \tag{C9} \label{C9}  \\
         g^{y_m y_m}\sqrt{g}= B_m \frac{\sqrt{f}}{\sqrt{c}} \frac{q_m y_m}{\prod_{i\not=m} q_i y_i} , \tag{C10} \label{C10}\\
         g^{y_m y_n}\sqrt{g}= \frac{1}{2} C_{mn} \frac{\sqrt{f}}{\sqrt{c}} \frac{1}{\prod_{i\not= m,i\not= n}q_i y_i} (m\not= n) . \tag{C11} \label{C11}
     \end{gather}
  Substituting Eq.\eqref{C9}, Eq.\eqref{C10}, Eq.\eqref{C11} into Eq.\eqref{C2}, we get the coefficients of the first-order terms. Comparing the coefficient of $\partial_{y_1}$ term of Eq.\eqref{C2} with that of Eq.(13), we have
     \begin{multline}
        B_1 q_1^2 y_1 + D_1 q_1 y_1 = B_1 q_1^2 y_1 + \frac{1}{2} B_1 (q_1 y_1)^2 \frac{\partial_{y_1} f}{f} \\
        +\sum_{k=2}^d \frac{1}{4} C_{1k} q_1 y_1 q_k y_k \frac{\partial_{y_k} f}{f} .
        \notag
     \end{multline}
   It can be simplified as
     \begin{equation}
         D_1=\frac{1}{2} B_1 q_1 y_1 \frac{\partial_{y_1} f}{f} +\frac{1}{4} C_{12} q_2 y_2 \frac{\partial_{y_2} f}{f} + \cdots + \frac{1}{4} C_{1d} q_d y_d \frac{\partial_{y_d} f}{f} .
         \notag
     \end{equation}
  Similarly, we take the same processes for $\partial_{y_2}, \cdots ,\partial_{y_d}$ terms and get
     \begin{equation}
         \begin{cases}
          D_1=\frac{1}{2} B_1 q_1 y_1 \frac{\partial_{y_1} f}{f} +\frac{1}{4} C_{12} q_2 y_2 \frac{\partial_{y_2} f}{f} + \cdots + \frac{1}{4} C_{1d} q_d y_d \frac{\partial_{y_d} f}{f} \\
          D_2=\frac{1}{4} C_{12} q_1 y_1 \frac{\partial_{y_1} f}{f} + \frac{1}{2} B_2 q_2 y_2 \frac{\partial_{y_2} f}{f} + \cdots + \frac{1}{4} C_{2d} q_d y_d \frac{\partial_{y_d} f}{f} \\
          \vdots  \\
          D_d= \frac{1}{4} C_{1d} q_1 y_1 \frac{\partial_{y_1} f}{f}+ \cdots + \frac{1}{2} B_d q_d y_d \frac{\partial_{y_d} f}{f}
         \end{cases}
         \tag{C12}
         \label{C12}
     \end{equation}
  We assume that the function $f$ has the form
     \begin{equation}
         f=\prod_{m=1}^d y_m^{\frac{A_m}{q_m}} ,
         \tag{C13}
         \label{C13}
     \end{equation}
  and then
     \begin{equation}
         q_i y_i \frac{\partial_{y_i} f}{f} =A_i .
         \tag{C14}
         \label{C14}
     \end{equation}
  Thus, the differential equation set Eq.\eqref{C12} is transformed to an algebraic equation set
     \begin{equation}
         \begin{cases}
          D_1=\frac{1}{2}  B_1 A_1 + \frac{1}{4} C_{12}  A_2 + \cdots + \frac{1}{4} C_{1d}  A_d \\
          D_2=\frac{1}{4} C_{12}  A_1 + \frac{1}{2} B_2  A_2 + \cdots + \frac{1}{4} C_{2d}  A_d\\
          \vdots  \\
          D_d= \frac{1}{4} C_{1d}  A_1 + \cdots + \frac{1}{2} B_d  A_d
         \end{cases}
         \tag{C15}
         \label{C15}
     \end{equation}
  Define $\mathbf{A}=(A_1, \cdots ,A_d)^T$ and $\mathbf{D}=(D_1, \cdots ,D_d)^T$. We can express Eq.\eqref{C15} as
     \begin{equation}
         \frac{1}{2} X \mathbf{A}=\mathbf{D} .
         \tag{C16}
         \label{C16}
     \end{equation}
  If $c\not= 0$, we can solve $\mathbf{A}$ directly. If $c=0$, we can change $B_1$ to $B_1+\delta$ and $\delta$ is a small parameter. In this case, $c\not= 0$ and we can get the solution $\mathbf{A}(\delta)$. And then, taking the $\delta \to 0$ limit, we can get $\mathbf{A}$.
  \par

  Then, comparing Eq.\eqref{C13} with Eq.\eqref{C6}, it is a natural choice to set
     \begin{equation}
         g_{z_i z_i}=y_i^{\frac{A_i}{q_i}} ,
         \tag{C17}
         \label{C17}
     \end{equation}
  and $\mathbf{gz}_d$ is determined analytically,
      \begin{equation}
          \mathbf{gz}_d=diag\{y_1^{\frac{A_1}{q_1}}, \cdots ,y_d^{\frac{A_d}{q_d}}\} .
          \tag{C18}
          \label{C18}
      \end{equation}
  If $d=1$, we have $g_{z_1 z_1} =y_1^{\frac{2 D_1}{B_1 q_1}}$, which is compatible with the results of the $1$-dimensional case given in Sec.$\mathbf{II}$.
  \par

  Combining Eq.\eqref{C18} and Eq.\eqref{C3}, the metric of the $2d$-dimensional curved space is obtained analytically, $\tilde{\mathbf{g}}_d= diag\{\mathbf{gy}_d, \mathbf{gz}_d\}$. Replacing $\{ q_1, \cdots ,q_d\}$ in $\tilde{\mathbf{g}}_d$ with $\{ w_1, \cdots ,w_d \}$, we obtain the metric of $\mathcal{A}_d$, i.e. $\mathbf{g}_d(w_1,w_2, \cdots, w_d)$.

\section{Special Cases}
 \subsection{$q_i=0$ cases}
    If $q_i=0$, the coordinate transformation $y_i=e^{q_i x_i}$ becomes invalid. In this case, let $y_i=x_i$, and then the terms about index $i$ in Eq.(13) become $B_i \partial_{y_i}^2$, $D_i \partial_{y_i}$ and $C_{ij} q_j y_j \partial_{y_i} \partial_{y_j}$($i\not= j$). By comparing the coefficients of Eq.(13) with those of Eq.\eqref{C2}, we find
      \begin{gather}
          g^{y_i y_i} =B_i  ,
          \tag{D1}  \label{D1}
          \\
          g^{y_i y_j} = \frac{1}{2} C_{ij} q_j y_j  = g^{y_j y_i} (i \not= j).
          \tag{D2}  \label{D2}
      \end{gather}
    \par

    Then, consider $\partial_{y_i}$ term. we have
     \begin{equation}
         g=\frac{f}{c} \prod_{m=1,m\not=i}^d (q_m y_m)^{-2} .
         \tag{D3}
         \label{D3}
     \end{equation}
    Thus, only the part of $\sqrt{f}$ contains variable $y_i$ in $g^{y_i y_i} \sqrt{g}$ and $g^{y_i y_j} \sqrt{g}$ ($i\not= j$). Comparing Eq.(13) with Eq.\eqref{C2}, we can find all the factors $q_i y_i$ in Eq.\eqref{C12} become $1$. If we take
     \begin{equation}
         \frac{\partial_{y_i} f}{f} = A_i ,
         \tag{D4}
         \label{D4}
     \end{equation}
    the function $f$ becomes
     \begin{equation}
         f=e^{A_i y_i} \prod_{m\not= i} y_m^{\frac{A_m}{q_m}} .
         \tag{D5}
         \label{D5}
     \end{equation}
    Hence, it is a natural choice to take $g_{z_i z_i} = e^{A_i y_i}$ in this case.
    \par

   If $i_1 \not= i_2 \not=  \cdots  \not= i_k$ ($k<d$), but $q_{i_1}=q_{i_2}= \cdots =q_{i_k}=0$, we take $y_{i_1}=x_{i_1}, \cdots ,y_{i_k}=x_{i_k}$. It is easy to find that all factors $q_{i_1} y_{i_1}, \cdots ,q_{i_k} y_{i_k}$ in $\mathbf{gy}_d^{-1}$ become $1$, and $g_{z_{i_1} z_{i_1}}=e^{A_{i_1} y_{i_1}}, \cdots ,g_{z_{i_k} z_{i_k}}=e^{A_{i_k} y_{i_k}}$.

  \subsection{The Hermitian limit}
    In the Hermitian limit, all states of $d$-dimensional HN model are extended. This means $q_{i,min}=q_{i,max}=0$ for $i=1,2, \cdots, d$. Since the Hamiltonian is Hermitian in this case, the hopping amplitudes must satisfy the relation
      \begin{equation}
          t_{j_1, \cdots ,j_i, \cdots ,j_d} = t_{j_1, \cdots ,-j_i, \cdots ,j_d} ,
          \tag{D6}
          \label{D6}
      \end{equation}
    for any $i$. Then, observing the definition of $D_i$ and $C_{mn}$ (Eq.\eqref{B7}, Eq.\eqref{B8}), we find all $D_i$ and $C_{mn}$ become $0$ in the Hermitian case. When $D_1=D_2= \cdots =0$, Eq.\eqref{C16} becomes
      \begin{equation}
          \frac{1}{2} X \mathbf{A} = \mathbf{0} .
          \tag{D7}
          \label{D7}
      \end{equation}
    Thus, we get $\mathbf{A}=\mathbf{0}$. Together with the results in Appendix C, we get
     \begin{gather}
         \mathbf{gy}_d^{-1} = diag\{ B_1, \cdots ,B_d \} ,\tag{D8}
         \label{D8}\\
         \mathbf{gz}_d = I_{d \times d} ,
         \tag{D9}
          \label{D9}
     \end{gather}
    and
      \begin{equation}
          \mathbf{g}_d(w_1, \cdots, w_d)=
             \begin{pmatrix}
                 \frac{1}{B_1} \\
                 & \frac{1}{B_2} \\
                 & & \ddots  \\
                 & & & \frac{1}{B_d} \\
                 & & & & 1 \\
                 & & & & & \ddots \\
                 & & & & & & 1
             \end{pmatrix}_{2d \times 2d},
             \tag{D10}
             \label{D10}
      \end{equation}
   in Hermitian case, where $w_1 = w_2 = \cdots = w_d = 0$. Since $B_i > 0$, $(i = 1, \cdots, d)$, $\mathbf{g}_d$ is a diagonal matrix, which means $\mathcal{A}_d$ is a flat manifold in the Hermitian case.

\section{Special example in $d>1$ cases}

    For high-dimensional non-Hermitian systems, non-Bloch band theory is invalid. But for some special cases of $d$-dimensional non-Hermitian HN model, in which Hamiltonian is separable about different coordinates, non-Bloch band theory is still valid. We can use $d$ generalized Brillouin zone(GBZ) to depict the continuous energy spectrum.
    \par

    In this case, the non-Hermitian Hamiltonian is
      \begin{equation}
          H=\sum_{x_1,x_2, \cdots, x_d} \sum_{p=1}^d \sum_{j_p = -l_p}^{r_p} t^{(p)}_{j_p} c_{x_1,x_2, \cdots, x_p+j_p, \cdots, x_d}^{\dagger} c_{x_1,x_2, \cdots, x_d},
          \label{E1}
      \end{equation}
    where $x_i \in [1,L_i]$ is the $i$th coordinate of the system. The Schr\"{o}dinger equation in real space is $H \Psi = E \Psi$. $\Psi$ is the wavefunction, i.e.
      \begin{equation}
          \Psi_{x_1,x_2, \cdots, x_d} = \sum_m \Phi^{(m)}_{x_1,x_2, \cdots, x_d},
          \label{E2}
      \end{equation}
    where $\Phi^{(m)}$ is the $m$th eigenvector of the bulk Schr\"{o}dinger equation.
    \par

    Similar to the $1$-dimensional non-Bloch band theory, we assume that
      \begin{equation}
          \Phi_{x_1,x_2, \cdots, x_d}= \beta_1^{x_1} \beta_2^{x_2} \cdots \beta_d^{x_d} \phi .
          \label{E3}
      \end{equation}
    The bulk Schr\"{o}dinger equation can be written as
      \begin{equation}
          \left[\sum_{p=1}^d \sum_{j_p=-l_p}^{r_p} t^{(p)}_{j_p} \beta_p^{-j_p}  \right] \phi = E \phi .
          \label{E4}
      \end{equation}
    We can decompose Eq.\eqref{E4} into $d$ independent equations
      \begin{equation}
        \left[\sum_{j_p=-l_p}^{r_p} t^{(p)}_{j_p} \beta_p^{-j_p} \right] \phi = E_p \phi
        \label{E5},
      \end{equation}
    for $p=1,2, \cdots, d$, and $\sum_{p=1}^d E_p =E$. To ensure that $E$ belongs to the continuous spectrum of the system, $E_p$ must be in the continuous spectrum of Eq.\eqref{E5} for each $p=1,2, \cdots, d$.
    \par

    We can obtain the characteristic equation about Eq.\eqref{E5}
      \begin{equation}
        \sum_{j_p=-l_p}^{r_p} t^{(p)}_{j_p} \beta_p^{-j_p} - E_p =0.
        \label{E6}
      \end{equation}
    It has $l_p + r_p$ roots $\beta_{p,1}, \beta_{p,2}, \cdots \beta_{p,l_p + r_p}$, and we let $|\beta_{p,1}| \leqslant |\beta_{p,2}| \leqslant \cdots \leqslant |\beta_{p,l_p + r_p}|$. Thus, for a decomposition $\{E_1, E_2, \cdots, E_d\}$ of $E$, the bulk Schr\"{o}dinger equation can give $\prod_p (l_p + r_p)$ eigenvectors, which are denoted by $\Phi^{(m_1,m_2, \cdots, m_d)}$ and
      \begin{equation}
          \Phi^{(m_1,m_2, \cdots, m_d)}_{x_1,x_2, \cdots, x_d} = \beta_{1,m_1}^{x_1} \beta_{2,m_2}^{x_2} \cdots \beta_{d,m_d}^{x_d} \phi^{(m_1,m_2, \cdots, m_d)},
          \label{E7}
      \end{equation}
    where, $m_p= 1,2,\cdots, l_p+r_p$ for $p=1,2,\cdots, d$. There are $d$ independent indexes $m_1,m_2, \cdots, m_d$ in $\phi^{(m_1,m_2, \cdots, m_d)}$. Hence, $\phi^{(m_1,m_2, \cdots, m_d)}$ can be rewritten as
      \begin{equation}
        \phi^{(m_1,m_2, \cdots, m_d)} = \phi_1^{m_1} \phi_2^{m_2} \cdots \phi_d^{m_d}.
        \label{E8}
      \end{equation}
    Substituting Eq.\eqref{E7} and Eq.\eqref{E8} into Eq.\eqref{E2}, we get the wavefunction in real space
      \begin{equation}
          \Psi_{x_1,x_2, \cdots, x_d} = \sum_{m_1,m_2, \cdots ,m_d} \prod_{p=1}^d \beta_{p,m_p}^{x_p} \phi_p^{m_p}.
          \label{E9}
      \end{equation}
    \par

    Now, we consider the boundary condition. The boundary condition about the first coordinate $x_1$ can be written as
      \begin{equation}
          \Psi_{-(r_1-1),x_2, \cdots, x_d} = \Psi_{-(r_1-2),x_2, \cdots, x_d} = \cdots = \Psi_{0,x_2, \cdots, x_d} = 0
          \label{E10}
      \end{equation}
      \begin{equation}
        \Psi_{L_1+1,x_2, \cdots, x_d} = \Psi_{L_1+2,x_2, \cdots, x_d} = \cdots = \Psi_{L_1+l_1,x_2, \cdots, x_d} = 0
        \label{E11}
    \end{equation}
    for arbitrary $x_i \in [1,L_i]$ ($i=2,3, \cdots, d$). Since Eq.\eqref{E10} and Eq.\eqref{E11} are independent of the values of $x_2, x_3, \cdots, x_d$, the two equations can be reduced as
      \begin{multline}
          \sum_{m_1} \beta_{1,m_1}^{-(r_1-1)} \phi_1^{m_1} = \sum_{m_1} \beta_{1,m_1}^{-(r_1-2)} \phi_1^{m_1} = \cdots \\
          = \sum_{m_1} \beta_{1,m_1}^0 \phi_1^{m_1} = 0
          \label{E12}
      \end{multline}
      \begin{multline}
        \sum_{m_1} \beta_{1,m_1}^{L_1+1} \phi_1^{m_1} = \sum_{m_1} \beta_{1,m_1}^{L_1+2} \phi_1^{m_1} = \cdots \\
        = \sum_{m_1} \beta_{1,m_1}^{L_1+l_1} \phi_1^{m_1} = 0.
        \label{E13}
    \end{multline}
    By combining Eq.\eqref{E12} and Eq.\eqref{E13}, we obtain the matrix form of the boundary condition
      \begin{equation}
          \mathcal{J}_1 \Phi_1 =0 \label{E14} ,
      \end{equation}
    where
      \begin{equation}
        \Phi_1 = (\phi_1^1, \phi_1^2, \cdots, \phi_1^{(l_1 + r_1)})^T  \label{E15}
      \end{equation}
    and
      \begin{equation}
        \mathcal{J}_1=
          \begin{pmatrix}
              \beta_{1,1}^{-(r_1-1)} & \beta_{1,2}^{-(r_1-1)} & \cdots & \beta_{1,l_1 + r_1}^{-(r_1-1)} \\
              \beta_{1,1}^{-(r_1-2)} & \beta_{1,2}^{-(r_1-2)} & \cdots &\beta_{1,l_1 + r_1}^{-(r_1-2)} \\
              \vdots & \vdots & \cdots &\vdots \\
              \beta_{1,1}^0 & \beta_{1,2}^0 & \cdots & \beta_{1,l_1 + r_1}^0 \\
              \beta_{1,1}^{1} \beta_{1,1}^{L1} & \beta_{1,2}^{1} \beta_{1,2}^{L1} & \cdots & \beta_{1,l_1 + r_1}^{1} \beta_{1,l_1 + r_1}^{L1}  \\
              \beta_{1,1}^{2} \beta_{1,1}^{L1} & \beta_{1,2}^{2} \beta_{1,2}^{L1} & \cdots & \beta_{1,l_1 + r_1}^{2} \beta_{1,l_1 + r_1}^{L1}  \\
              \vdots & \vdots & \cdots & \vdots \\
              \beta_{1,1}^{l_1} \beta_{1,1}^{L1} & \beta_{1,2}^{l_1} \beta_{1,2}^{L1} & \cdots & \beta_{1,l_1 + r_1}^{l_1} \beta_{1,l_1 + r_1}^{L1}
          \end{pmatrix} .
          \label{E16}
      \end{equation}
    There exists nontrival solutions of $\Phi_1$ only if $ det(\mathcal{J}_1) = 0$.
    \par

     Similar to the $1$-dimensional non-Bloch band theory, if $E_1$ belongs to the continuous spectrum of Eq.\eqref{E5} for $p=1$ case, the absolute value of $\beta_{1,r_1}$ and $\beta_{1,r_1+1}$ must be equal, i.e. $|\beta_{1,r_1}| = |\beta_{1,r_1+1}|$ (Ref.\cite{19}). This gives a curve in the complex plane denoted by $GBZ_1$. When $\beta_1 \in GBZ_1$, $E_1$ can be obtained by Eq.\eqref{E5}, which can change continuously with the variation of $\beta_1$ in $GBZ_1$.
    \par

    By using the same method, we can get other $d-1$ curves $GBZ_2, GBZ_3, \cdots, GBZ_d$. The value of $E_i$ can be given through Eq.\eqref{E5} over the $GBZ_i$($i=1,2, \cdots, d$).

    The eigenenergy $E$ of this non-Hermitian system can be obtained by $E= \sum_{p=1}^d E_p$, which belongs to the continuous spectrum determined by
      \begin{equation}
          E(\beta_1,\beta_2,\cdots, \beta_d) = \sum_{p=1}^d \sum_{j_p=-l_p}^{r_p} t^{(p)}_{j_p} \beta_p^{-j_p} ,
          \label{E17}
      \end{equation}
    where $\beta_i \in GBZ_i$ ($i=1,2,\cdots,d$). In other word, non-Bloch band theory is valid for these cases.

\section{Taylor Expension of $\phi(x)$}
    In this appendix, we elucidate why we keep the Taylor expansion up to the quadratic term in Sec.$\mathbf{II}$.
    \par

    Since we consider the case that $\phi(x)$ changes smoothly with the variation of position, The contribution of high-order terms is small in the Taylor expansion. However, it is not appropriate for just keeping the first order term. 
    \par 
    
    If we truncate the Taylor expansion at the first order term, $\phi(x+j) = \phi(x) + j \partial_x \phi(x)$. Substituting it into Eq.(3), we get
      \begin{equation}
          C \partial_x \phi(x) = A \phi(x),
          \label{F1}
      \end{equation}
    where $C$ and $A$ are the same in Sec.$\mathbf{II}$. Under Hermitian limit, $t_j = t_{-j}$ and $q=0$. Hence, $C=0$ and Eq.\eqref{F1} become
      \begin{equation}
          A \phi(x) = 0
          \label{F2} .
      \end{equation}
    The solution of Eq.\eqref{F2} is $\phi(x) = 0$. However, the wavefunction under Hermitian case is plane wave. Thus, truncating the Taylor expansion at the first order term is not a good approximation. If we remain the quadratic term, we get Eq.(4) and the solution is 
      \begin{equation}
          \phi(x) = \alpha e^{\lambda x + i\delta},
          \label{F3}
      \end{equation}
    where $\alpha , \delta \in \mathbb{R}$ are constants and
      \begin{equation}
          \lambda =  \frac{-C \pm \sqrt{C^2 + 4 B A}}{2 B}
          \label{F4}.
      \end{equation}
    Under Hermitian limit $C =0$, $B > 0$ and $A = E - \sum_j t_j = E - t_0 - \sum_{j>0} 2 t_j$. For Hermitian system, the energy spectrum can be given by $E(k) = t_0 + \sum_{j>0} 2 t_j cos(jk)$ and $E_{max} = t_0 + \sum_{j>0} 2 t_j$. Thus, under Hermitian limit, $A \leqslant 0$, and Eq.\eqref{F4} becomes
      \begin{equation}
          \lambda = \pm i \sqrt{\frac{|A|}{B}}
          \label{F5} .
      \end{equation}
    Apparently, $\lambda$ is a pure imaginary number and $\phi(x)$ return to a pure plane wave naturally and continuously under Hermitian limit.
    \par

    Under the condition $\frac{2\pi}{max\{ \theta_1, \theta_2\} } >> (j_{max}-j_{min})$, the frequency of the vibration of the phase~($e^{-i \theta x}$) is very low. Thus, the contribution of the higher order terms must be small and keeping the quadratic term is good enough to retain the information about the vibration of the phase.

    \par

    Hence, we think truncating the Taylor expansion at the quadratic term is a good approximation if $|\theta|$ is small.

\section{The Validity of The Generalization}

    We start from Eq.(3), which is the Schr\"{o}dinger equation on lattice. Under continuous position variable $x$, the wavefunction $\Phi = (\phi_1, \phi_2, \cdots, \phi_L)$ is replaced by function $\phi(x)$, which satisfies $\phi(x)|_{x=n} = \phi_n$, and $\psi(x) = e^{-q x} \phi(x)$. Applying Taylor expansion to $\phi(x)$ at $x=n$, we have
      \begin{equation}
          \phi(x-j) = \phi(x) + \sum_{p=1}^{\infty} \frac{(-j)^p}{p!} \partial_x^p \phi(x) \label{G1}.
      \end{equation}  
    Substituting it into Eq.(3), we get
      \begin{equation}
          \sum_{p=1}^{\infty} \alpha_p \partial_x^p \phi(x) = A \phi(x) \label{G2}
      \end{equation}
    where
      \begin{equation}
          \alpha_p = \sum_j \frac{(-j)^p}{p!} t_j e^{qj} , \label{G3}
      \end{equation}
    and $A = E - \sum_j t_j e^{qj}$. Observing Eq.\eqref{G2} and Eq.\eqref{G3}, we find that
      \begin{equation}
          \phi(x) = \frac{1}{\sqrt{L}} e^{- i \theta x} \label{G4}
      \end{equation}
    is the solution of the Schr\"{o}dinger equation Eq.(G2) with eigenenergy
      \begin{equation}
          E = \sum_j t_j e^{qj} e^{i \theta j}.  \label{G5} 
      \end{equation}
    Substituting Eq.\eqref{G4},\eqref{G5} into Eq.\eqref{G2} and expanding the factor $e^{i \theta j}$, we get
      \begin{equation}
        \sum_{p=1}^{\infty} \alpha_p \partial_x^p \phi(x) = \sum_{p=1}^{\infty} \sum_j t_j e^{qj} \frac{(i \theta j)^p}{p!} \phi(x) \label{G6}
      \end{equation}
    \par

    If $|\theta|$ is small, the contribution of the high order terms in the left hand side~(LHS) of Eq.\eqref{G2} is very small. Hence, we can only retain the first and the second order term in the LHS of Eq.\eqref{G2} and get Eq.(4)~($\alpha_1 = C$, $\alpha_2 = B$). This is consistent with the conclusion in Appendix F.
    \par

    If $|\theta|$ is finite, the contribution of the high order terms in the LHS of Eq.\eqref{G2} cannot be neglect. Thus, Eq.(4) is no longer a good approximation in this case. 
    \par

    Observing Eq.\eqref{G6}, we find that there exist an one-to-one correspondence between the terms in the left hand side~(LHS) and right hand side~(RHS) of this equation. In other word, for wavefunction $\phi(x)=\frac{1}{\sqrt{L}} e^{-i \theta x}$, 
      \begin{equation}
        \alpha_p \partial_x^p \phi(x) = \sum_j t_j e^{qj} \frac{(i \theta j)^p}{p!} \phi(x). \label{G7}
      \end{equation}
    Thus, we introduce 
      \begin{equation}
          \tilde{A}(q,\theta) = \sum_{p=1}^{2} \sum_j t_j e^{qj} \frac{(i \theta j)^p}{p!} . \label{G8}
      \end{equation}
    We find that $\phi(x)$ always satisfies the equation
      \begin{equation}
          B \partial_x^2 \phi(x) + C \partial_x \phi(x) = \tilde{A} \phi(x) ,\label{G9}
      \end{equation} 
    for $\theta \in [0,2\pi]$. Thus, for the case $|\theta|$ is finite, we need to use Eq.\eqref{G9} to establish the correspondence. The difference between Eq.(4) and Eq.\eqref{G9} are the coefficients $A$ and $\tilde{A}$. However, the metric  obtained~(Eq.(9)) under the case $|\theta|$ is small does not contain $A$. Hence, the metric of corresponding curved space for the case $|\theta|$ is finite will still be Eq.(9).   
    \par

    Eq.\eqref{G9} has two solutions,
      \begin{equation}
          \phi_1 (x) = \frac{1}{\sqrt{L}} e^{\lambda_1 x} \qquad
          \phi_2 (x) = \frac{1}{\sqrt{L}} e^{\lambda_2 x} ,
          \label{H1}
      \end{equation}
    with 
       \begin{equation}
           \begin{aligned}
            \lambda_1 =  \frac{-C + \sqrt{C^2 + 4 B \tilde{A}}}{2 B} \\
            \lambda_2 =  \frac{-C - \sqrt{C^2 + 4 B \tilde{A}}}{2 B}.
           \end{aligned}
           \label{H2}
       \end{equation}
    Based on Eq.\eqref{G8},(17) and (18), 
       \begin{equation}
           \tilde{A}(q, \theta) = - i \theta C - \theta^2 B. \label{H3}
       \end{equation}
    Substituting Eq,\eqref{H3} into Eq.\eqref{H2}, we get
       \begin{equation}
           \lambda_1 = - i \theta  \qquad  \lambda_2 = -\frac{C}{B} + i \theta .
           \label{H4}
       \end{equation}
    Thus, 
       \begin{gather}
           \phi_1(x) = \frac{1}{\sqrt{L}} e^{-i \theta x} \label{H5} \\
           \phi_2(x) = \frac{1}{\sqrt{L}} e^{-\frac{C}{B} x + i \theta x}. \label{H6}
       \end{gather}
     Eq.\eqref{G9} is reduced from the original Schr\"{o}dinger equation, Eq.\eqref{G2}. $\phi_1(x)$ satisfies Eq.(G2) and $\phi_2(x)$ does not satisfy Eq.(G2). Hence, the localized solution, $\phi_2$ is not the exact solution of the original Schr\"{o}dinger equation and we do not consider it.
    \par

    Now, we consider the open boundary condition. Eq.\eqref{H5} satisfies both Eq.\eqref{G2} and Eq.\eqref{G9}, but does not satisfy OBC. And solutions of Eq.\eqref{G2} and Eq.\eqref{G9} under OBC are different. However, there exist an one-to-one correspondence between them.
    \par

    According to non-Bloch band theory, for a $1$-dimensional non-Hermitian HN model with finite length $L$, we can choose $\beta_{m,1} = e^{-q_m - i \theta_{m,1}} \in GBZ$, $\beta_{m,2} = e^{-q_m - i \theta_{m,2}} \in GBZ$ for $m=1,2,\cdots, L$, such that $E(\beta_{m,1}) = \sum_j t_j \beta_{m,1}^{-j} = E(\beta_{m,2}) = E_m$ with $\theta_{m,1} - \theta_{m,2} = \frac{2 \pi m}{L}$. Thus for $A_m = E_m - \sum_j t_j e^{q_m j}$, there are two solutions
       \begin{equation}
           \phi_{m,1}(x) = \frac{1}{\sqrt{L}} e^{-i \theta_{m,1} x} \qquad
           \phi_{m,2}(x) = \frac{1}{\sqrt{L}} e^{-i \theta_{m,2} x},
           \label{H7}
       \end{equation}
    which satisfy Eq.\eqref{G2} with $A_m$. Then, considering the wavefunction
       \begin{equation}
           \tilde{\phi} _m(x) = \frac{1}{\sqrt{2}} (\phi_{m,1}(x) - \phi_{m,2}(x)), \label{H8}
       \end{equation}
    we find $\tilde{\phi} _m(x)$ satisfies Eq.(G2) and the open boundary condition~($\tilde{\phi} _m(0) = \tilde{\phi} _m(L) = 0$) with parameters $q_m$, $B$, $C$, $\theta_{m,1}$ and eigenenergy $E_m$. When $L \rightarrow \infty$, $\{ E_m \}$ gives the spectrum of this system under the thermodynamic limit. 
    \par
      
    $\tilde{\phi} _m$ is not the solution of Eq.\eqref{G9} under OBC since $\tilde{A}(q_m,\theta_{m,1} ) \not= \tilde{A}(q_m,\theta_{m,2} )$ although $A(q_m, \theta_{m,1} ) = A(q_m, \theta_{m,2} )$. Solving Eq.\eqref{G9} under OBC, we can get a solution
      \begin{equation}
          \tilde{\tilde{\phi}}_m(x) = \frac{1}{s} e^{\frac{-C}{2 B} x} (e^{i \theta_{m,1} x} - e^{-i \theta_{m,1} x}) \label{H9}
      \end{equation}
      with parameters $q_m$, $B$, $C$, $\theta_{m,1}$ and eigenvalue 
        \begin{equation}
            \tilde{\tilde{A}}_m(q_m,B,C, \theta_{m,1}) = \frac{-C^2 - \theta_{m,1}^2}{ 4 B} \label{H10},
        \end{equation}
    where, $s$ is a constant for normalization. Thus, we obtain the one-to-one correspondence $\tilde{\phi}_m^{ (q_m,B,C,\theta_{m,1})} (x)\Leftrightarrow \tilde{\tilde{\phi}}_m^{(q_m,B,C,\theta_{m,1})} (x)$ and $A_m(q_m,B,C,\theta_{m,1}) \Leftrightarrow  \tilde{\tilde{A}}_m (q_m,B,C,\theta_{m,1})$. 
    \par

    Since the Schr\"{o}dinger equation in curved space~(Eq.(7)) with metric $\mathbf{g}_1$~(Eq.(9)) can be transformed to Eq.(G9) by a coordinate transformation $x= \frac{1}{q} ln y$, the wavefunction in curved space with eigenenergy 
      \begin{multline}
          E^{\prime} (q_m,B,C,\theta_{m,1}) = -\frac{\hbar^2}{2 M} \tilde{\tilde{A}}_m (q_m,B,C,\theta_{m,1})  \\
          = -\frac{\hbar^2}{2 M} \frac{-C^2 - \theta_{m,1}^2}{ 4 B} \label{H11}
      \end{multline}

      is $\phi^{\prime (q_m,B,C,\theta_{m,1})}(y) = \tilde{\tilde{\phi}}_m^{(q_m,B,C,\theta_{m,1})} (x(y))$. Thus, we get the correspondence relations about the wavefunctions and eigenenergies of non-Hermitian HN model in flat space and free particle system in curved space,
        \begin{align}
            \tilde{\phi}_m^{ (q_m,B,C,\theta_{m,1})} (x) \Leftrightarrow \phi^{\prime (q_m,B,C,\theta_{m,1})}(y) \label{H12}   \\
            A_m(q_m,B,C,\theta_{m,1}) \Leftrightarrow E^{\prime} (q_m,B,C,\theta_{m,1}) \label{H13}.
        \end{align}

    \par

    It is similar for high dimensional case. Expanding Eq.\eqref{B4}, which is the Schr\"{o}dinger equation about Hamiltonian Eq.(10), we have
      \begin{equation}
          \sum_{p=1}^{\infty} \hat{\eta}_p \phi(x_1,\cdots,x_d) = K \phi(x_1,\cdots,x_d) \label{G10},
      \end{equation}
    where 
      \begin{equation}
          K= E - \sum_{j_1, \cdots ,j_d} t_{j_1, \cdots ,j_d} (\prod_{i=1}^d e^{q_i j_i}) \label{G11}
      \end{equation}
    and $\hat{\eta}_p$ is an operator defined as
      \begin{equation}
          \hat{\eta}_p = \sum_{j_1, \cdots ,j_d} t_{j_1, \cdots ,j_d} (\prod_{i=1}^d e^{q_i j_i}) \sum_{m_1, \cdots, m_p = 1}^d \frac{\prod_{i=1}^p -j_{m_i} \partial_{x_{m_i}}}{p!}   . \label{G12}
      \end{equation}
    We can find that
      \begin{equation}
        \phi(x_1,\cdots,x_d) = \frac{1}{\sqrt{V}} e^{-i \sum_{m=1}^d \theta_m x_m} \label{G13}
      \end{equation}
    is the solution of Eq.\eqref{G10} with $V=\prod_{i=1}^d L_i$ and 
      \begin{equation}
          E = \sum_{j_1, \cdots ,j_d} t_{j_1, \cdots ,j_d} (\prod_{m=1}^d e^{q_m j_m} e^{i \theta_m j_m}). \label{G14}
      \end{equation}
    Substituting Eq.\eqref{G12},\eqref{G13},\eqref{G14} into Eq.\eqref{G10} and expanding the factor $\prod_{m=1}^d e^{i \theta_m j_m}$, we find there is also an one-to-one correspondence between the LHS and the RHS of Eq.\eqref{G10},
      \begin{equation}
          \hat{\eta}_p \phi(x_1,\cdots,x_d) = \sum_{j_1, \cdots ,j_d} t_{j_1, \cdots ,j_d} (\prod_{m=1}^d e^{q_m j_m}) \frac{1}{p!} (\sum_{m=1}^d i \theta_m j_m)^p. \label{G15}
      \end{equation}
      \par

    If $|\theta|_i$ is small for all $i=1,2,\cdots,d$, the contribution of high order terms in the LHS of Eq.(G10) is small. Hence, we can just keep the first and the second term and get Eq.\eqref{B10}. Once some $|\theta|_i$ are finit, Eq.(B10) will no longer be a good approximation of Eq.\eqref{G10}. Similar to the $1$-dimensional case, we introduce
      \begin{multline}
          \tilde{K}(q_1,\cdots, q_d, \theta_1, \cdots, \theta_d) = \sum_{p=1}^2 \sum_{j_1, \cdots ,j_d} t_{j_1, \cdots ,j_d} \\
          \times (\prod_{m=1}^d e^{q_m j_m}) \frac{1}{p!} (\sum_{m=1}^d i \theta_m j_m)^p. \label{G16}
      \end{multline}
    Then, $\phi(x_1,\cdots,x_d)$ here always satisfies
      \begin{equation}
          \hat{\eta}_2 \phi(x_1,\cdots,x_d) + \hat{\eta}_1 \phi(x_1,\cdots,x_d) = \tilde{K} \phi(x_1,\cdots,x_d). \label{G17}
      \end{equation}
    Considering the definition of $\hat{\eta}_p$~(Eq.\eqref{G12}) and the definition of $B_m$, $D_m$, $C_{mn}$ in Appendix B~(Eq.\eqref{B6},\eqref{B7},\eqref{B8}), Eq.\eqref{G17} can be rewritten as
      \begin{equation}
        \left[\sum_{i=1}^d B_i \partial_{x_i}^2 + \sum_{i=1}^d D_i \partial_{x_i} + \sum_{m<n} C_{mn} \partial_{x_m} \partial_{x_n} - \tilde{K} \right]\phi = 0 . \label{G18}
      \end{equation}
    If some $|\theta|_i$ are finite, we need to use Eq.\eqref{G18} to construct the correspondence but not Eq.\eqref{B10}. Comparing Eq.\eqref{G18} with Eq.\eqref{B10}, the difference between them are $A$ and $-\tilde{K}$. However, the metric of the $2d$-dimensional curved space for all $|\theta|_i$ are small does not contain $A$~(Eq.(15),(16)). This means for the case some $|\theta_i|$ are finite, the metric of corresponding curved space can still be given by Eq.(14),(15),(16).  Since non-Bloch band theory is invalid for high-dimensional case, we can not get the concrete correspondence relations about the wavefunctions and eigenenergies between the two systems for general case.

   \bibliography{paper}
\end{document}